\documentclass[preprint,12pt]{elsarticle}

\usepackage{geometry}
\usepackage{graphics,graphicx}
\usepackage[usenames,dvipsnames]{xcolor}
\usepackage{amsmath}
\usepackage{latexsym}
\usepackage{amsfonts}
\usepackage{mathrsfs}
\usepackage{bigints}
\usepackage{bm}
\usepackage{multirow}
\usepackage{amsthm}
\usepackage{amssymb}
\usepackage{caption}
\usepackage{subcaption}
\usepackage{float}
\usepackage[colorlinks,citecolor=blue,urlcolor=blue,bookmarks=false,hypertexnames=true]{hyperref}
\usepackage[title]{appendix}
\usepackage{tikz}
\usetikzlibrary{shapes.geometric, arrows}
\tikzstyle{startstop} = [rectangle, rounded corners, minimum width=3cm, minimum height=1cm,text centered, draw=black]
\tikzstyle{arrow} = [thick,->,>=stealth]
\graphicspath{ {Fig/} }

\usepackage[labelformat=simple]{subcaption}

\newtheorem{definition}{Definition}[section]
\newtheorem{theorem}{Theorem}[section]

\newtheorem{lemma}[theorem]{Lemma}

\usepackage{tikz}

\usetikzlibrary{shapes.geometric, arrows, shapes}
\tikzstyle{startstop} = [rectangle, rounded corners, minimum width=3cm, minimum height=1cm,text centered, draw=black]
\tikzstyle{arrow} = [thick,->,>=stealth]

\newcommand\solidrule[1][1cm]{\rule[0.6mm]{6mm}{2pt}}

\newcommand\dashdotrule{\mbox{%
\rule[0.5mm]{2.5mm}{2pt}\hspace{1mm}\rule[0.5mm]{1mm}{2pt}\hspace{1mm}\rule[0.5mm]{2.5mm}{2pt}}}

\definecolor{fadblu}{rgb}{0, 0.4470, 0.7410}
\definecolor{darkmar}{rgb}{0.6350, 0.0780, 0.1840}
\definecolor{mag}{rgb}{1.00,0.07,0.65}

\DeclareRobustCommand{\legendsquare}[1]{%
  \textcolor{#1}{\rule{2.5ex}{1.5ex}}%
}
\definecolor{fadblu1}{rgb}{0.2, 0.40, 0.76}
\definecolor{mag1}{rgb}{0.8,0.07,0.7}
\definecolor{gray1}{rgb}{0.5,0.5,0.5}

\definecolor{GRN}{rgb}{0.4660, 0.6740, 0.1880}
\definecolor{MRN}{rgb}{0.6350, 0.0780, 0.1840}

\biboptions{sort&compress}

\makeatletter

\newcommand{\mathleft}{\@fleqntrue\@mathmargin0pt}
\newcommand{\mathcenter}{\@fleqnfalse}
\makeatother
\newlength{\bibitemsep}
\newlength{\bibparskip}
\let\oldthebibliography\thebibliography
\renewcommand\thebibliography[1]{%
  \oldthebibliography{#1}%
  \setlength{\parskip}{\bibitemsep}%
  \setlength{\itemsep}{\bibparskip}%
}

\usetikzlibrary{positioning, fit, calc}   
\tikzset{block/.style={draw, thick, text width=2cm ,minimum height=1.3cm, align=center},   
line/.style={-latex}     
}

\begin{document}

\begin{frontmatter}


\title{
Innate behavioural mechanisms and defensive traits in ecological models of predator-prey types
}

\author[inst1]{Sangeeta Saha\corref{cor1}}
\ead{sangeetasaha629@gmail.com}
\author[inst1]{Swadesh Pal}
\ead{spal@wlu.ca}
\author[inst1,inst2]{Roderick Melnik}
\ead{rmelnik@wlu.ca}

\cortext[cor1]{Corresponding author}
\address[inst1]{MS2 Discovery Interdisciplinary Research Institute, Wilfrid Laurier University, Waterloo, Canada}
\address[inst2]{BCAM - Basque Center for Applied Mathematics, E-48009, Bilbao, Spain}

\begin{abstract}
There are various examples of phenotypic plasticity in ecosystems that serve as the basis for a wide range of inducible defences against predation. These strategies include camouflage, burrowing, mimicry, evasive actions, and even counterattacks that enhance survival under fluctuating predatory threats. Additionally, the ability to exhibit plastic responses often influences ecological balances, shaping predator-prey coexistence over time. This study introduces a predator-prey model where prey species show inducible defences, providing new insights into the role of adaptive strategies in these complex interactions. Moreover, the predator's consumption rate is assumed to be influenced by mutual interference, so the Beddington-DeAngelis response is chosen for this work. The stabilizing impact of the defensive mechanism is one of several intriguing outcomes produced by the dynamics. Moreover, the predator population rises when the interference rate increases to a moderate value even in the presence of lower prey defence but decreases monotonically for stronger defence levels. Furthermore, we identify a bistable domain when the handling rate is used as a control parameter, emphasizing the critical role of initial population sizes in determining system outcomes. By considering the species diffusion in a bounded region, the study is expanded into a spatio-temporal model. The numerical simulation reveals that the Turing domain decreases as the level of protection increases. The study is subsequently extended to incorporate taxis, known as the directed movement of species toward or away from another species. Our investigation identifies the conditions under which pattern formation emerges, driven by the interplay of inducible defences, taxis as well as species diffusion. Numerical simulations demonstrate that including taxis within the spatio-temporal model exerts a stabilizing influence, thereby diminishing the potential for pattern formation in the system.

\end{abstract}


\end{frontmatter}


\section{Introduction} \label{sec:1}

Studies in ecological and evolutionary dynamics have shown that ecological and evolutionary processes can occur simultaneously \cite{hairston2005rapid, schoener2011newest}. There is existing literature that highlights the importance of phenotypic plasticity-based intraspecific variation in species ecology and evolution \cite{gangur2017changes}. Baldwin, an American psychologist, demonstrated how phenotypic plasticity promotes adaptive evolution and survival \cite{baldwin1896new}. The phrase ``phenotypic plasticity" defines how an organism brings changes in behaviour, morphology, and physiology when put in a certain environment \cite{yamamichi2019modelling}. The predator-prey system, which serves as a model system for population dynamics and is the foundation of the complicated food chain, and food network, has long been a major issue in ecology and biomathematics \cite{wu2016global, arsie2022predator, saha2023analysis, saha2024nonlocal}. The predator-prey system's phenotypic plasticity has a profound influence on species evolution. Phenotypic plasticity, which occurs naturally, includes induced adaptations against predation like as concealment, cave-dwelling, mimicry, evasion, counterattacking, and so on. Inducible defences are an essential ecological component that influences ecological dynamics, particularly predator-prey stability, either directly or indirectly. 

Researchers have been studying that inducible defences have a major impact on predator-prey dynamics \cite{van2018inducible}, e.g., size-selective predator \cite{riessen2009turning}. The idea was then confirmed using a model system with inducible defences: the water flea Daphnia's induction of a neck spine in reaction to the phantom midge Chaoborus's predatory larvae. The inducible defences of Daphnia were also investigated by other researchers using experiments \cite{boeing2010inducible, rabus2011growing}. The differences in the protective spine length of rotifers were examined via field and laboratory research \cite{zhang2017predator}. Anuran tadpole and predatory dragonfly nymph were used in lab experiments by several researchers to investigate how prey uses chemical or visual cues to avoid predators \cite{gomez2011invasive, takahara2012inducible}. Experimentally demonstrated that exposure to predators strengthens anti-predator responses in susceptible animals \cite{west2018predator}. To properly study the mechanism of inducible defences, theoretical methods must be applied to biological phenomena that experimentation cannot explain, such as fitness gradient, optimal trait, and switching function \cite{yamamichi2019modelling}. There are bi-trophic and tri-trophic food chain models incorporating consumer-induced polymorphisms to evaluate how inducible defences affect community stability and persistence in the presence of other factors \cite{vos2004binducible, hammill2010predator, liu2023dynamics, saha2024role}.

The Holling type II functional response was chosen by Ramos-Jiliberto et al. in their work \cite{ramos2007pre} as follows:
$$\phi_{i}=\frac{1}{H_{i}+(A_{i}N_{i-1})^{-1}}$$
with $A_{i}$ and $H_{i}$ as the attack rate and the handling time of a prey item, respectively, and after implementing the inducible defence factor, it takes the form as (parameters are given in the article): 
$$\phi_{y}=\frac{1}{h_{y}[1+(E_{y}-1)D]+\{a_{y}[1+(F_{y}-1)D]x\}^{-1}}.$$
In their work, they have observed the system's qualitative characteristics. \par

Gonz$\Acute{a}$lez-Olivares et al. \cite{gonzalez2017multiple} put out a conceptual model to classify the different kinds of prey defences. There, it is thought that the prey's defensive traits include changes to its appearance, behaviour, physiology, or life history in response to predator pressure. Conversely, the rate of consumption of the predators will be lower in the presence of the non-consumptive impacts of the top predators \cite{gonzalez2017multiple}. With $U$ and $V$ representing the prey and predator biomass, respectively, the functional response has the form $p(U)=qU(1-R)$. With $U_{r}$ representing the biomass of predator-resistant prey, the parameter $R=U_{r}/U$ represents the defensive trait set. The scientists classified prey defences into six types based on $R$'s sensitivity to prey and predator biomass. In addition, they have taken into account the fact that $\partial R/\partial U=0$ and $\partial R/\partial V>0$ hold, indicating that the average prey's resistance to predators depends on the biomass of predators. We call these kinds of defensive responses ``inducible defences." \par 

The functional response is a crucial element in depicting the species' behavioural characteristics. The dynamic behaviour of the predator is more significantly influenced by how the predator consumes their prey. Most complex dynamical behaviours, including chaotic states, periodic oscillations, stable states, etc., are dictated by functional responses. The prey biomass, the predators' efficiency in locating, capturing, and killing the prey, their competition, and other variables all influence the functional responses. Holling-type and Lotka–Volterra–type prey-dependent functional responses are the most commonly employed. But in these functional responses, it is assumed that the predators do not interfere with one another's activities \cite{holling1959components}. This explains why the Holling type II functional response's per capita predation rate is an increasing, saturating, and smooth function of prey density. The concept of predator-density-dependent functional responses emerges because prey-dependent functional responses cannot adequately characterise predator interference. The functional response in a predator-prey framework should be predator-dependent in many situations, especially when predators must search for food (and share or compete for it). Numerous studies have suggested that predator reliance in the functional response is prevalent in natural and lab environments \cite{arditi1990underestimation, dolman1995intensity}. Predators disrupt one other's activities to produce competitive effects, and several studies and observations suggest that prey alters its behaviour due to increased predator danger. Territorial disputes, an adverse habitat, or insufficient prey biomass can all contribute to this interference. For this reason, models with a predator-dependent functional response can be a good alternative to models with a prey-dependent functional response \cite{skalski2001functional}. Beddington \cite{beddington1975mutual} and DeAngelis et al. \cite{deangelis1975model} independently proposed a functional response in 1975 that took into account predators' mutual interference to mediate between theoretical and experimental viewpoints \cite{huisman1997formal}. 


Relationships between living things and their natural environments are the only focus of ecological research. Ecological systems are developed by their interactions, which are important for population ecology. In nature, resources are frequently distributed extensively over the ecosystem. Therefore, organisms are diffusive throughout the environment because they must locate food and survive. Since trophic interactions are greatly influenced by species movement, different spatial patterns in nature have evolved. Several studies have demonstrated that aquatic and terrestrial populations may form patterns \cite{chang2016spatial, wang2004stationary}. Spatial patterns are shaped by many causes, including deterministic processes, species growth, mobility, stochastic processes, environmental changes, and more. Because species interact across habitats, patterns are common in the ecological system \cite{melese2011pattern, pal2023role}. Self-organized spatial patterns are the result of a deterministic process in interacting organisms. The spatio-temporal predator-prey model is given as follows:
\mathcenter
\begin{equation}\label{eq:int1}
\begin{aligned} 
\frac{\partial U}{\partial T} &= D_{U}\Delta U+RU\left(1-\frac{U}{K}\right)-f(U,V)V, \\
\frac{\partial V}{\partial T} &= D_{V}\Delta V+\sigma_{1}f(U,V)V-\delta_{1}V,
\end{aligned}
\end{equation}
where $U(\Tilde{\mathbf{x}},T)$ and $V(\Tilde{\mathbf{x}},T)$ represent the densities of the prey species and predator species
at spatial location $\Tilde{\mathbf{x}}$ and time $T$, respectively, and $f(U, V)$ are chosen as different kinds of functional responses depending on the corresponding analysis. The intrinsic growth rate of the prey and its carrying capacity in the environment are denoted by the parameters $R$ and $K$. The natural mortality rate of the predator species is represented by the parameter $\delta_{1}$, whereas the biomass conversion rate is $\sigma_{1}$. The prey and predator species' relative self-diffusion is indicated by the parameters $D_{U}$ and $D_{V}$. The random movement of species from places with higher concentrations to areas with lower concentrations is known as the diffusion strategy. \par

In this article, we concentrated on a predator-prey system where the prey species choose an inducible defence strategy towards their predator. Moreover, we have chosen the Beddington-DeAngelis response, a predator-dependent functional response, where it is considered that the predator species spend some time encountering each other except searching for and processing the prey. So, in system \eqref{eq:int1}, the functional response is going to be as follows: 
$$f(U,V)=\frac{M(1-s(V))U}{1+HM(1-s(V))U+\Gamma V}$$
with $s(V)=\Lambda V/(\Phi+V)$. Here, $M$ is the predator's consumption rate, $\Lambda$ is the rate of defence by the prey population, and $\Phi$ is a positive constant. The parameter $H$ is the time required for the predator to handle a prey individual, and $\Gamma$ is the per capita interference rate among predators. In the Beddington–DeAngelis functional response, the interference occurs only during the searching process for prey, and so, the interference effect decreases as prey densities increase. Now, the reaction-diffusion system, which implements the inducible defence in the presence of predator interference, has the form
\mathcenter
\begin{equation}\label{eq:int2}
\begin{aligned} 
\frac{\partial U(\Tilde{\mathbf{x}},T)}{\partial T} &= D_{U}\Delta U+RU\left(1-\frac{U}{K}\right)-\frac{M(1-s(V))U}{1+H(1-s(V))U+\Gamma V}, \\
\frac{\partial V(\Tilde{\mathbf{x}},T)}{\partial T} &= D_{V}\Delta V+\frac{\sigma_{1}M(1-s(V))U}{1+HM(1-s(V))U+\Gamma V}-\delta_{1}V,
\end{aligned}
\end{equation}
defined in a bounded domain $\Omega\subset \mathbb{R}^{n}$ with smooth boundary and outer normal $\nu$, supplemented with initial conditions: $U(\Tilde{\mathbf{x}},0)=U_{0}\geq 0, \ V(\Tilde{\mathbf{x}},0)=V_{0}\geq 0,\ \Tilde{\mathbf{x}}\in \Omega,$ and homogeneous Neumann boundary conditions: $\partial U/\partial\nu=0,\ \partial V/\partial\nu=0$, on $\Tilde{\mathbf{x}}\in \partial\Omega, \ t>0$.

It is noted that (\ref{eq:int2}) is a standard reaction-diffusion system, where the prey and the predators obey the random movements due to the terms $\Delta U$ and $\Delta V$. As it is well known, random movements describing the moving process of prey and predator species may not be practical. Indeed, the movement of prey and predator species may spontaneously follow directional movement in the real world. For example, the predators will move to a higher prey density for food. On the contrary, the prey will move in the predator’s opposite direction for survival. Therefore, such phenomena should be manifested in the model.

Now, taxis are directed movements of species in response to specific stimuli, such as phototaxis in the presence of light  \cite{foster1980light, ghorai2005penetrative} or chemotaxis in the presence of chemical gradients \cite{lin1988large, friedman2002stability, issa2019persistence}. Taxis allow organisms to migrate towards or away from certain stimuli. In mathematical modelling, two types of directed taxis are recognized in population biology: prey-taxis, where the predator follows the prey \cite{lee2009pattern, ma2012stationary, jin2017global, wang2018boundedness, xiang2018global, han2024spatio, peng2024dynamic, zhang2024predator}, and predator-taxis, during which the prey flees from the predator \cite{wu2018dynamics, wang2017pattern, dai2020global}. Kareiva and Odell \cite{kareiva1987swarms} proposed the first ecological model involving the prey-taxis to describe the directed movements of species. Since then, various ecological models that describe the directed movement of the species have been proposed and investigated. Taxis-based motions significantly influence spatio-temporal pattern formation \cite{maini1991bifurcating, sleeman2005existence, myerscough1998pattern}. For instance, a reaction-diffusion model with Rosenzweig-MacArthur kinetics is unable to generate Turing structures \cite{banerjee2011self} but may produce stationary Turing patterns when the concept of taxis is implemented \cite{sapoukhina2003role, zhang2019pattern}. Bell and Haskell studied a predator-prey system with both direct and indirect taxis mechanisms to establish the global existence of positive solutions and have shown the existence of non-trivial steady-state solutions \cite{bell2021attraction}. The global existence and uniform boundedness of classical solutions of a predator-prey system with the predator-taxis were studied in \cite{ahn2021global}, and it was shown that large chemosensitivity could give rise to the occurrence of pattern formations. The influence of the fear and predator-taxis was considered in \cite{dong2022influence}, where it is found that there is a single Hopf bifurcation point when the fear level was chosen as the bifurcation parameter, but the system admits two different Hopf bifurcation points when predator-taxis sensitivity is taken as the bifurcation parameter. Also, Wang et al., in their work, have shown that both attractive prey-taxis and repulsive predator-taxis compress the spatial patterns, while repulsive prey-taxis and attractive predator-taxis help to generate spatial patterns \cite{wang2021pattern}. Inspired by these works, we have tried to explore the prey- as well as predator-taxis into the system (\ref{eq:int2}). \par

This study integrates defence-induced prey- and predator-taxis movement of species into the corresponding spatio-temporal system of \eqref{eq:int2}. Many plants and animals produce a variety of chemicals (pheromones, kairomones, and so on) that are employed for both inter- and intraspecific interaction. For example, olfaction is a fundamental mechanism by which prey detect predators and respond to anti-predator signals. There are other forms of anti-predator responses too that defend against predation; however, we have focused on the inducible defensive mechanism used by prey species when attacked by predators. In \eqref{eq:int3}, we have analyzed the impact of prey- and predator-taxis on the system. In this situation, the prey species attempt to avoid (or get away) areas where they detect high predator populations. Prey avoidance of predators is influenced by a variety of factors (e.g. predator scent and predatory calls), which promote species movements outward along the predator gradient. As a defensive measure, the prey species travels not just at random but also away from the predator species' high-density zones. The predator's hunting policy, on the other hand, combines random dispersion with directed movement towards the gradient of prey (chemicals released from prey harmed during an attack). The predator looks for prey not just in a random direction (diffusion), but also in locations with high prey density. So, we propose the reaction-diffusion-advection system as follows: 
\mathcenter
\begin{equation}\label{eq:int3}
\begin{aligned} 
\frac{\partial U(\Tilde{\mathbf{x}},T)}{\partial T} &= D_{U}\Delta U+\nabla\cdot(\xi_{0}U\nabla V)+RU\left(1-\frac{U}{K}\right)-\frac{M(1-s(V))U}{1+HM(1-s(V))U+\Gamma V}, \\
\frac{\partial V(\Tilde{\mathbf{x}},T)}{\partial T} &= D_{V}\Delta V-\nabla\cdot(\eta_{0} V\nabla U)+\frac{\sigma_{1}M(1-s(V))U}{1+HM(1-s(V))U+\Gamma V}-\delta_{1}V.
\end{aligned}
\end{equation}
with non-negative initial conditions and no-flux boundary conditions on $\Omega$. Based on real-world application concerns, we select a two-dimensional bounded spatial domain $(\mbox{i.e.,}\ n = 2)$. So, $\Tilde{\mathbf{x}}=(\Tilde{x},\Tilde{y})$ denotes the two-dimensional spatial domain $\Omega$ defined by $0<\Tilde{x}<\Tilde{L}$ and $0<\Tilde{y}<\Tilde{L}$. The phrase $\nabla\cdot(\xi_{0}U\nabla V)$ reflects predator-taxis, which indicates the tendency of prey moving toward the opposite direction of the increasing gradient of predators owing to anti-predator behaviour. $\xi_{0}$ is the sensitivity coefficient, which assesses the intensity of prey species' directed movement farther from predators, so, the predator-taxis is called repulsive and attractive when $\xi_{0}>0$ and $\xi_{0}<0$, respectively. On the other hand, the term $-\nabla\cdot(\eta_{0}V\nabla U)$ symbolizes prey-taxis, which indicates the tendency of a predator moving toward the direction of the gradient of prey density function to get more effectiveness in hunting. $\eta_{0}$ is the sensitivity coefficient, which quantifies predator species' directed movement towards prey species, and hence, the prey-taxis is called attractive and repulsive when $\eta_{0}>0$ and $\eta_{0}<0$, respectively. In this work, we have chosen repulsive predator-taxis $(\xi_{0}>0)$ and attractive prey-taxis $(\eta_{0}>0)$ to delineate the scenario where predators are moving towards high prey density and the preys are avoiding the high predator density by exhibiting inducible defence. Dealing with a non-dimensional version of the model \eqref{eq:int3} can simplify the study by reducing the number of parameters and thus, considering the non-dimensional species, space, and time variables as
$$U=Ku,\ T=\frac{t}{R},\ V=RKv,\ \tilde{\mathbf{x}}=\frac{1}{\sqrt{R}}\mathbf{x}.$$
Substituting the non-dimensional variables into the system (\ref{eq:int3}) and dropping the tildes on the dimensionless variables $U, V, T$, and $\tilde{\mathbf{x}}$, we then obtain:
\mathcenter
\begin{equation}\label{eq:diff3}
\begin{aligned} 
\frac{\partial u(\mathbf{x},t)}{\partial t} &= d_{1}\Delta u+\nabla\cdot(\xi u\nabla v)+u(1-u)-\frac{f(v)uv}{m+hf(v)u+\gamma v}, \\
\frac{\partial v(\mathbf{x},t)}{\partial t} &= d_{2}\Delta v-\nabla\cdot(\eta v\nabla u)+\frac{\sigma f(v)uv}{m+hf(v)u+\gamma v}-\delta v,
\end{aligned}
\end{equation}
where $f(v)=1-\omega v/(\phi+v)$ with the initial conditions $u(\cdot,0)=u_{0}\geq 0, \ \ v(\cdot,0)=v_{0}\geq 0$ and boundary conditions $\langle{\nabla u, \nu}\rangle,\ \langle{\nabla v, \nu}\rangle=0$ on $\partial\Omega, \ t>0$, and the new dimensionless parameters to be the following:
\begin{align*}
\omega&=\Lambda,\ \phi=\frac{\Phi}{RK}, \ m=\frac{1}{KM}, \ h=H, \ \gamma=\frac{\Gamma R}{M}, \ \xi=\xi_{0}RK, \ d_{1}=D_{U},\\ 
\sigma&=\frac{\sigma_{1}}{R},\ \delta=\frac{\delta_{1}}{R},\ \eta=\eta_{0}K, \ d_{2}=D_{V}.
\end{align*}
We have considered the parametric restriction $0\leq \omega<1$ to maintain biological relevance. 

In system \eqref{eq:diff3}, the dynamic behaviour of a predator-prey system is examined when prey demonstrates inducible defence in the presence of predator interference. Both prey and predator-taxis are induced in the system due to these two behavioural features. However, we have yet to find any study that investigates the impact of this sort of defensive mechanism in the presence of predator-dependent functional response. This is likely the first study to look at defence-induced taxis in an ecological model. In the following sections, we have primarily shown the biological well-definedness of the spatio-temporal model, the local stability of the homogeneous steady states, the existence and non-existence of heterogeneous steady states, and conditions for pattern formation.


\section{The temporal model} \label{sec:2}

We first analyze the dynamics of the predator-prey interaction when the prey-taxis, as well as the predator-taxis, is not implemented (put $\xi=\eta=0$ in \eqref{eq:diff3}). The model, in the absence of population diffusion, takes the form
\begin{equation}\label{eq:det1}
\begin{aligned}
\frac{du}{dt}&=u(1-u)-\frac{f(v)uv}{m+hf(v)u+\gamma v},\ \ u(0)\geq 0 \\
\frac{dv}{dt}&=\frac{\sigma f(v)uv}{m+hf(v)u+\gamma v}-\delta v,\ \ v(0)\geq 0.
\end{aligned}  
\end{equation} 
The system parameters are positive and are the same as those described for model \eqref{eq:diff3}. Moreover, the parametric restriction $0\leq \omega<1$ holds. 

\subsection{Positivity and boundedness}\label{subsec-2.1}
\noindent The following theorem here shows that the model (\ref{eq:det1}) is well-posed as the solutions, when they exist, are positive, unique, and bounded.

\begin{theorem}
Solutions of system (\ref{eq:det1}), starting in $\mathbb{R}_{+}^{2}$, are positive and uniformly bounded for $t>0$.
\end{theorem}
\begin{proof}
Functions on the right-hand side of the system (\ref{eq:det1}) are continuous and locally Lipschitzian (as they are polynomials and rationals in $(u,v)$), so there exists a unique solution $(u(t),v(t))$ of the system with positive initial conditions $(u(0), v(0)) \geq 0$ on $[0,\tau],$ where $0<\tau<+\infty$ \cite{hale1977}. 
$$\textrm{Let}\ \psi_{1}(u,v)=(1-u)-\frac{f(v)v}{m+hf(v)u+\gamma v}\ \textrm{and}\ \psi_{2}(u,v)=\frac{\sigma f(v)u}{m+hf(v)u+\gamma v}-\delta.$$
From the first and second equation of (\ref{eq:det1}), we have
\begin{equation*}
u(t)=u(0)\exp\left[\int^t_0 \psi_{1}(u(z),v(z))\,dz\right]\ \textrm{and}\ v(t)=v(0)\exp\left[\int^t_0 \psi_{2}(u(z),v(z))\,dz\right].
\end{equation*}
So, the solutions of the system (\ref{eq:det1}) are feasible with time, i.e., $u(t)\geq 0$ and $v(t)\geq 0$ when $u(0)\geq 0$ and $v(0)\geq 0$. 

Now, we prove the solutions, starting in $\mathbb{R}_{+}^{2}$, are uniformly bounded too for $t>0$.
The first equation of (\ref{eq:det1}) gives:

\begin{equation*}
\begin{aligned}
\frac{du}{dt}&=u(1-u)-\frac{f(v)uv}{m+hf(v)u+\gamma v}\leq u(1-u) \\
\displaystyle \Rightarrow & \limsup_{t\rightarrow \infty}u(t)\leq 1.
\end{aligned}
\end{equation*}

\noindent Consider, $W(t)= u(t)+\frac{1}{\sigma}v(t)$. Then we have
\begin{equation*}
\begin{aligned}
\frac{dW}{dt}= \left(\frac{du}{dt}+\frac{1}{\sigma}\frac{dv}{dt}\right)&=[u(1-u)+u]-u-\frac{\delta v}{\sigma} \leq 1-\tau W, \ \ \ \mbox{where}\ \tau=\min\{1,\delta\}.
\end{aligned}
 \end{equation*}
Now, using the comparison theorem on the above differential inequality, we have
$$\displaystyle 0\leq W(t)\leq \frac{1}{\tau}+ \left(W(0)-\frac{1}{\tau}\right)\exp(-\tau t),$$ where $W(0)=W(u(0),v(0))$. As $\displaystyle t\rightarrow\infty,\ 0<W(t)\leq 1/\tau+\epsilon$ for sufficiently small $\epsilon>0$. Henceforth, the solutions of system (\ref{eq:det1}), initiating from the positive initial condition, enter into the region: 
$$\mathcal{T} = \{(u,v)\in \mathbb{R}^{2}_{+}: 0<u(t)\leq1; 0\leq W(t)\leq 1/\tau+\epsilon, \epsilon>0\}.$$
\end{proof}

\subsection{Equilibrium analysis}\label{subsec-2.2}
Solving the prey- and predator-nullcline, we obtain that the temporal system has:
\begin{enumerate}[(i)]
    \item The trivial equilibrium point $E_{0}=(0,0)$;
    \item The predator-free equilibrium point $E_{1}=(1,0)$; 
    \item One or more interior equilibrium points $E^{*}=(u^{*},v^{*})$, where $\displaystyle u^{*}= \delta(m+\gamma v^{*})/[(\sigma-h\delta)f(v^{*})]$ and $v^{*}$ satisfies the following equation: 
\begin{align}\label{eq:2.2}
   g(v)\equiv A_{1}v^{4}+A_{2}v^{3}+A_{3}v^{2}+A_{4}v+A_{5}=0,
\end{align}
where $A_{1}= \sigma\delta\gamma^{2}, \ A_{5}=\sigma m\phi^{2}[\delta(m+h)-\sigma], \\
A_{2}= (\sigma-h\delta)^{2}(1-\omega)^{2}-\sigma\gamma(1-\omega)(\sigma-h\delta)+2\sigma\delta\gamma(m+\gamma\phi), \\
A_{3}=2\phi(1-\omega)(\sigma-h\delta)^{2}-\sigma(\sigma-h\delta)\{m(1-\omega)+\gamma\phi(2-\omega)\}+\sigma\delta[(m+\gamma\phi)^{2}+2m\phi\gamma], \\ 
A_{4}= \phi^{2}(\sigma-h\delta)^{2}-\sigma\phi(\sigma-h\delta)\{\gamma\phi+m(2-\omega)\}+2m\sigma\phi\delta(m+\gamma\phi)$.
\end{enumerate}
Now, $g(0)=A_{5}<0$ when $\sigma>\delta(m+h)$, and $\lim_{v\rightarrow\infty}g(v)=+\infty$. So, there will be at least one positive root of the equation for $\sigma>\delta(m+h)$. Moreover, the feasibility of $u^{*}$ also holds when the mentioned condition is satisfied. It means the system (\ref{eq:det1}) contains at least one feasible interior equilibrium point $E^{*}$ when the parametric restriction is fulfilled.

\subsection{Local stability analysis}\label{subsec-2.3}
The local stability criterion of the equilibrium points can be determined by analyzing the eigenvalues of corresponding Jacobian matrices. The Jacobian matrix of system (\ref{eq:det1}) at an equilibrium point $(u,v)$ is 
\mathcenter
\begin{equation}\ \label{eq:det2}
\textbf{J}=  \begin{pmatrix}
a_{11} & a_{12} \\
a_{21} & a_{22} 
\end{pmatrix},
\end{equation}
where $\displaystyle a_{11}=1-2u-(m+\gamma v)f(v)vZ,\ a_{12}=-[mu(f(v)+f'(v))+hu^{2}f^{2}(v)+\gamma uv^{2}f'(v)]Z,  \\
a_{21}=\sigma f(v)v(m+\gamma v)Z,\ a_{22}=\sigma Z[mu(f(v)+f'(v))+hu^{2}f^{2}(v)+\gamma uv^{2}f'(v)]-\delta$, where $Z=1/[m+hf(v)u+\gamma v]^{2}$.

The Jacobian matrix at $E_{0}=(0,0)$ gives the eigenvalues as $1$ and $-\delta$. As one of the eigenvalues is positive, $E_{0}$ is a saddle point. Again, for the predator-free equilibrium $E_{1}=(1,0),$ the Jacobian matrix admits two eigenvalues as $\lambda_{1}=-1<0$ and $\lambda_{2}=\sigma/(m+h)-\delta$. So, $\lambda_{2}<0$ when $\sigma<\delta(m+h)$ holds. Hence, $E_{1}$ is locally asymptotically stable (LAS) when the above-mentioned condition is satisfied. Otherwise, this equilibrium acts as a saddle point. 
Now, the Jacobian matrix at $E^{*}$ is obtained by substituting $u=u^{*}$ and $v=v^{*}$ at $\mathbf{J}$ in \eqref{eq:det2} and the characteristic equation corresponding to $\textbf{J}(E^{*})$ is given as follows: 
\mathcenter
\begin{equation}\label{eq:2.5}
\lambda^{2}+C_{1}\lambda+C_{2}=0,
\end{equation}
where $C_{1}=-\mbox{tr}(\textbf{J}(E^{*}))=-(a_{11}+a_{22})$ and $C_{2}=\det(\textbf{J}(E^{*}))=a_{11}a_{22}-a_{12}a_{21}$. To satisfy the local stability condition of Routh-Hurwitz criteria, $C_{1}>0$ and $C_{2}>0$ need to be fulfilled, which implies
\begin{equation} \label{eq:2.6}
    a_{11}+a_{22}<0\ \textrm{and}\ a_{11}a_{22}>a_{12}a_{21}
\end{equation}
have to be held.

\subsection{Local bifurcations around the equilibrium points}\label{subsec-2.4}
The local bifurcations around the equilibrium points are analyzed mainly with the help of Sotomayor's theorem and Hopf's bifurcation theorem \cite{perko2013differential}. In the system, if the stability condition of any of the equilibrium points violates in such a way that the corresponding determinant becomes $0$, giving a simple zero eigenvalue, then there will occur transcritical bifurcation, and we can observe the exchange of stability in that bifurcation threshold. The following theorem states the condition where such bifurcation can be observed in $E_{1}$, but before proceeding further, we define $\overline{\textbf{F}}=(\overline{F}_{1},\overline{F}_{2})^{T},$ where 
\begin{equation}\label{eq:10}
 \overline{F}_{1}(u,v)=u(1-u)-\frac{f(v)uv}{m+hf(v)u+\gamma v},\ \
 \overline{F}_{2}(u,v)=\frac{\sigma f(v)uv}{m+hf(v)u+\gamma v}-\delta v.    
\end{equation}

\begin{theorem}
System (\ref{eq:det1}) shows a transcritical bifurcation around $E_{1}(1,0)$ at $\displaystyle h_{tc}=\sigma/\delta-m$, choosing $h$ as the bifurcating parameter.
\end{theorem}
\begin{proof}
From the Jacobian matrix corresponding to $E_{1}$, the eigenvalues are given by $\lambda_{1}=-1$ and $\lambda_{2}=\sigma/(m+h)-\delta$, which are negative by the stability condition. Let $h_{tc}$ be the value of $h$ such that $\sigma=\delta(m+h)$ so that $\textbf{J}|_{E_{1}}$ has a simple zero eigenvalue at $h_{tc}$. So, at $h=h_{tc}:$
\begin{equation*}
\textbf{J}|_{E_{1}}=\begin{pmatrix}
-1 & -1/(m+h) \\
0 & 0
\end{pmatrix}.
\end{equation*}Now, the calculations give the eigenvectors corresponding to the zero eigenvalues of $\textbf{J}|_{E_{1}}$ and $\textbf{J}|^{T}_{E_{1}}$ at $h=h_{tc}$ as $\textbf{V}=(-1, m+h)^{T}$ and $\textbf{W}=(0,1)^{T}$, respectively. Therefore, we have
\begin{equation*}
\begin{aligned}
\Omega_{1}&= \textbf{W}^{T}\cdot\overline{\textbf{F}}_{h}(E_{1}, h_{tc})=\frac{-\sigma f^{2}(v)u^{2}v}{[m+hf(v)u+\gamma v]^{2}}\bigg|_{E_{1}}=0, \\
\Omega_{2}&= \textbf{W}^{T}\left[D\overline{\textbf{F}}_{h}(E_{1}, h_{tc})\textbf{V}\right]=-\delta\neq0, \\
\textrm{and}\ \Omega_{3}&= \textbf{W}^{T}\left[D^{2}\overline{\textbf{F}}(E_{1},h_{tc})(\textbf{V},\textbf{V})\right]=\frac{2}{\sigma}[m\delta(\gamma-\sigma)-\sigma\gamma] \neq0. \\
\end{aligned}
\end{equation*}
\mathcenter
Hence, by Sotomayor's theorem, the system undergoes a transcritical bifurcation around $E_{1}$ at $h=h_{tc}$.
\end{proof}

If any of the mentioned inequalities in (\ref{eq:2.6}) is violated, then the equilibrium point becomes unstable, and the system performs oscillatory or non-oscillatory behaviour. In fact, the system starts to oscillate around $(u^{*}, v^{*})$ if $C_{1}>0$ along with $C_{1}^{2}-4C_{2}<0$ as the eigenvalues will be in the form of the complex conjugate in this case. So, we get the following theorem.

\begin{theorem}
If $E^{*}$ exists with the feasibility conditions, then a simple Hopf bifurcation occurs at unique $\omega=\omega_{H}$, where $\omega_{H}$ is the positive root of $C_{1}(\omega)=0$, providing $C_{2}(\omega_{H})>0$ (stated in equation (\ref{eq:2.5})).
\end{theorem}
\begin{proof}
At $\omega=\omega_{H}$, the characteristic equation of system (\ref{eq:det1}) at $E^{*}$ is $(\omega^{2}+C_{2})=0$. So, the equation has a pair of purely imaginary roots $\lambda_{1}=i\sqrt{C_{2}}$ and $\lambda_{2}=-i\sqrt{C_{2}}$, where $C_{2}(\omega)$ is a continuous function of $m$. Now, in a small neighbourhood of $\omega_{H},$ the roots are $\lambda_{1}=p_{1}(\omega)+ip_{2}(\omega)$ and $\lambda_{2}=p_{1}(\omega)-ip_{2}(\omega)$ ($p_{1}$ and $p_{2}$ are $\mathcal{C}^{1}$ functions in $\mathbb{R}$). \\
To show the transversality condition, we check $\displaystyle \frac{d}{d\omega}[Re(\lambda_{i}(\omega))]\bigg|_{\omega=\omega_{H}}\neq 0,$ for $i=1,2.$ \\
Put $\lambda(\omega)=p_{1}(\omega)+ip_{2}(\omega)$ in (\ref{eq:2.5}), we get
\mathcenter
\begin{equation} \label{eq:2.7}
(p_{1}+ip_{2})^{2}+C_{1}(p_{1}+ip_{2})+C_{2}=0.
\end{equation}
Differentiating (\ref{eq:2.7}) with respect to $\omega$, we get
\begin{equation*}
2(p_{1}+ip_{2})(\dot{p_{1}}+i\dot{{p_{2}}})+C_{1}(\dot{p_{1}}+i\dot{p_{2}})+\dot{C_{1}}(p_{1}+ip_{2})+\dot{C_{2}}=0.
\end{equation*}
Comparing the real and imaginary parts from both sides, we have 
\begin{subequations}\label{eq:2.8}
\begin{align} 
(2p_{1}+C_{1})\dot{p_{1}}-(2p_{2})\dot{p_{2}}+(\dot{C_{1}}p_{1}+\dot{C_{2}}) = 0, \label{eq:2.8a} \\
(2p_{2})\dot{p_{1}}+(2p_{1}+C_{1})\dot{p_{2}}+\dot{C_{1}}p_{2}=0. \label{eq:2.8b}
\end{align}
\end{subequations}
Solving \eqref{eq:2.8a} and \eqref{eq:2.8b} we get, $\displaystyle \dot{p_{1}}=\frac{-2p_{2}^{2}\dot{C_{1}}-(2p_{1}+C_{1})(\dot{C_{1}}p_{1}+\dot{C_{2}})}{(2p_{1}+C_{1})^{2}+4p_{2}^{2}}$. \\
At, $p_{1}=0,\ p_{2}=\pm \sqrt{C_{2}}:\ \displaystyle \dot{p_{1}}=\frac{-2\dot{C_{1}}C_{2}-C_{1}\dot{C_{2}}}{C_{1}^{2}+4C_{2}}\neq 0$. Hence, this completes the proof.
\end{proof}


\section{The spatio-temporal model} \label{sec:3}
System \eqref{eq:diff3} considers the fact that the populations are heterogeneously distributed in the environment in the presence of taxis and hence, it depends not only on time but also on spatial positions. In this section, we have explored the dynamical nature of the system in the absence of attraction-repulsion taxis in a bounded domain $\Omega\subset \mathbb{R}^{2}$ with closed boundary $\partial\Omega$ and $\overline{\Omega}=\Omega\cup \partial\Omega$. The model \eqref{eq:diff3}, in absence of taxis, becomes
\begin{equation}\label{eq:diff2}
    \begin{aligned} 
\frac{\partial u}{\partial t}&=d_{1}\Delta u+u(1-u)-\frac{f(v)uv}{m+hf(v)u+\gamma v}, \\
\frac{\partial v}{\partial t}&=d_{2}\Delta v+\frac{\sigma f(v)uv}{m+hf(v)u+\gamma v}-\delta v,
\end{aligned}
\end{equation}
where $f(v)=1-\omega v/(\phi+v)$ with non-negative initial conditions and no-flux (Neumann) boundary conditions. As a two-dimensional spatial domain is chosen for the species movement, then we have $\Delta\equiv\partial^{2}/\partial x^{2}+\partial^{2}/\partial y^{2}$.

\subsection{Invariance and uniform persistence}
In this subsection, the invariance and uniform persistence of positive
steady-state $(u^{*}, v^{*})$ are studied for the diffusive predator-prey system \eqref{eq:diff2} under the above-mentioned initial and boundary conditions. First, we will show that any
non-negative solution $(u(\mathbf{x}, t), v(\mathbf{x}, t))$ of \eqref{eq:diff2} lies in a certain bounded region as $t\rightarrow\infty$ for all $x\in \Omega$.
In this regard, let us denote $\mathcal{D}=\Omega\times(0,\infty),\ \partial\mathcal{D}=\partial\Omega\times(0,\infty)$ and $\overline{\mathcal{D}}=\overline{\Omega}\times(0,\infty)$ where $\overline{\Omega}=\Omega\cup \partial\Omega$. 
To study the existence of a positively invariant attracting region, 
the boundedness and the persistence property of solutions of the spatio-temporal system (\ref{eq:diff2}), the following lemma is used \cite{wang2008global}.

\begin{lemma}\label{lemma-3.1}
Let f(s) be a positive $\mathcal{C}^{1}$ function for $s\geq 0$, and let $d>0$, $ \beta\geq 0$ be constants. Further, let $T\in[0,\infty)$ and $w\in \mathcal{C}^{2,1}(\Omega\times(T,\infty))\cap \mathcal{C}^{1,0}(\overline{\Omega}\times[T,\infty))$ be a positive function and satisfies
\begin{equation*}
 \begin{aligned} 
        w_{t}-d\Delta w&\leq (\geq)~w^{1+\beta}f(w)(\alpha-w) ~~ \mbox{in}\ \Omega\times [T,\infty), \\
        \frac{\partial w}{\partial\nu} &=0  ~~ \mbox{on}\ \partial\Omega\times [T,\infty), \\
    \end{aligned}
    \end{equation*}
for some constant $\alpha>0$, then 
$$\displaystyle \limsup_{t\rightarrow\infty}\max_{ \overline{\Omega}}w(\cdot,t)\leq \alpha \ \left(\liminf_{t\rightarrow\infty}\min_{\overline{\Omega}}w(\cdot,t)\geq \alpha\right).$$

\end{lemma}

\begin{theorem} \label{theorem-3.2}
All solutions of (\ref{eq:diff2}) are non-negative. Moreover, the The non-negative and non-trivial solutions $(u(\mathbf{x},t),v(\mathbf{x},t))$ of the system (\ref{eq:diff2}) satisfy
$$\displaystyle \limsup_{t\rightarrow\infty}\max_{\mathbf{x}\in \overline{\Omega}}u(\mathbf{x},t)\leq 1 \ \mbox{and}\ \limsup_{t\rightarrow\infty}\max_{\mathbf{x}\in \overline{\Omega}}v(\mathbf{x},t)\leq \frac{(\sigma-m\delta)}{\gamma\delta}.$$
\end{theorem}

\begin{proof}
Since the initial conditions of the system are non-negative, then the strong maximum principle gives $u(\mathbf{x},t)>0$ and $v(\mathbf{x},t)>0$. Using the positivity of $u$ and $v$, we find 
$$u_{t}-d_{1}\Delta u=u(1-u)-\frac{f(v)uv}{m+hf(v)u+\gamma v}\leq u(1-u).$$
Applying lemma \ref{lemma-3.1}, we obtain $\displaystyle\limsup_{t\rightarrow\infty}\max_{\mathbf{x}\in \overline{\Omega}}u(\mathbf{x},t)\leq 1=G_{u}$ (say). Then, for an arbitrary $\epsilon>0$, there exists $T>0$ such that $u(\mathbf{x},t)\leq 1+\epsilon$ in $\Omega\times[T,\infty)$. On the other hand, the second equation of (\ref{eq:diff2}) gives
$$v_{t}-d_{2}\Delta v=\frac{\sigma f(v)uv}{m+hf(v)u+\gamma v}-\delta v\leq v\left[\frac{\sigma u}{m+\gamma v}-\delta\right]\leq \left(\frac{\gamma\delta}{m}\right)v\left[\frac{\sigma-m\delta}{\gamma\delta}-v\right],$$
we obtain $\displaystyle\limsup_{t\rightarrow\infty}\max_{\mathbf{x}\in \overline{\Omega}}v(\mathbf{x},t)\leq \left(\frac{\sigma-m\delta}{\gamma\delta}\right)=G_{v}$ (say), provided $\sigma>m\delta$. Hence, the theorem is proved.   
\end{proof}

\begin{definition}
The system (\ref{eq:diff2}) is said to be persistent if for any non-negative initial data $(u_{0}(\mathbf{x}),v_{0}(\mathbf{x}))$ there exists a positive constant $\epsilon_{0}\equiv \epsilon_{0}(u_{0},v_{0})$ such that the solution $(u(\mathbf{x},t),v(\mathbf{x},t))$ of (\ref{eq:diff2}) satisfies the following conditions
$$\displaystyle \liminf_{t\rightarrow\infty}\min_{\mathbf{x}\in \overline{\Omega}}u(\mathbf{x},t)\geq \epsilon_{0}, \ \liminf_{t\rightarrow\infty}\min_{\mathbf{x}\in \overline{\Omega}}v(\mathbf{x},t)\geq \epsilon_{0}.$$
\end{definition}

\begin{theorem}\label{theorem-3.3}
The solutions of the system (\ref{eq:diff2}) are persistent for $$\gamma>\max\left\{1, \frac{\sigma\phi\delta+(\sigma-m\delta)^{2}}{\phi\delta[\sigma-\delta(m+h)]}\right\}\ \mbox{with}\  \sigma\neq\delta(m+h).$$
\end{theorem}

\begin{proof}
From the first equation of the system (\ref{eq:diff2}) we can write
\begin{align*}
u_{t}-d_{1}\Delta u= u(1-u)-\frac{f(v)uv}{m+hf(v)u+\gamma v}\geq u(1-u)-\frac{u}{\gamma}=u\left[\left(1-\frac{1}{\gamma}\right)-u\right].
\end{align*}
Then by lemma \ref{lemma-3.1}, we have $\displaystyle \liminf_{t\rightarrow\infty}\min_{\mathbf{x}\in \overline{\Omega}}u(\mathbf{x},t)\geq 1-1/\gamma=g_{u}$ (say). Therefore, for any $\epsilon>0$ with $0<\epsilon<g_{u}$ there exists a $T>0$ such that
$$u(\mathbf{x},t)\geq g_{u}-\epsilon \ \mbox{in}\ \overline{\Omega}\times[T, \infty).$$
Now, from the second equation, we can write 
\begin{align*}
v_{t}-d_{2}\Delta v&=v\left[\frac{\sigma[\phi+(1-\omega)v]u}{(m+\gamma v)(\phi+v)+hu[\phi+(1-\omega)v]}-\delta\right] \\
&\geq v\left[\frac{\sigma\phi g_{u}}{\phi(m+h)+\gamma G_{v}^{2}+v[h(1-\omega)+(m+\gamma\phi)]}-\delta\right] \\
&\geq \frac{v(\sigma\phi g_{u}-\delta B_{1})}{B_{1}+B_{2}G_{v}}\left[1-\frac{\delta B_{2}v}{(\sigma\phi g_{u}-\delta B_{1})}\right],
\end{align*}
where $B_{1}=\phi(m+h)+\gamma G_{v}^{2}$ and $B_{2}=(m+\gamma\phi)+h(1-\omega)$. Lemma \ref{lemma-3.1} gives $$\displaystyle \liminf_{t\rightarrow\infty}\min_{\mathbf{x}\in \overline{\Omega}}v(\mathbf{x},t)\geq \left(\frac{\sigma\phi g_{u}-\delta B_{1}}{\delta B_{2}}\right)=g_{v}\ (\mbox{say}).$$ 

\noindent Considering $\epsilon_{0}=\min\{g_{u}, g_{v}\}$, we obtain
$\displaystyle \liminf_{t\rightarrow\infty}\min_{\mathbf{x}\in \overline{\Omega}}u(\mathbf{x},t)\geq \epsilon_{0}, \ \liminf_{t\rightarrow\infty}\min_{\mathbf{x}\in \overline{\Omega}}v(\mathbf{x},t)\geq \epsilon_{0}.$
\end{proof}

\subsection{Diffusion-driven instability of the spatio-temporal model}

In this section, we intend to find the Turing bifurcation conditions around the homogeneous equilibrium $E^{*}=(u^{*},v^{*})$ of the diffusive predator-prey model \eqref{eq:diff2}. If the homogeneous steady state of the temporal model is locally stable to infinitesimal perturbation but becomes unstable in the presence of diffusion, a scenario of Turing instability occurs. 

\begin{theorem}\label{Theorem-3.5}
Suppose the temporal system has a positive interior equilibrium point $E^{*}$ which is locally asymptotically stable. Then, the spatio-temporal model \eqref{eq:diff2} does not possess Turing instability if $d_{2}a_{11}+d_{1}a_{22}<2\sqrt{d_{1}d_{2}(a_{11}a_{22}-a_{12}a_{21})}$ holds and the positive constant steady state is always locally asymptotically stable.
\end{theorem}

For the temporal model, $E^{*}=(u^{*},v^{*})$ is locally asymptotically stable when $C_{1}=-(a_{11}+a_{22})>0$ and $C_{2}=(a_{11}a_{22}-a_{12}a_{21})>0$ hold. Here, we apply heterogeneous perturbation around $E^{*}$ to obtain the criterion for instability of the spatio-temporal model. For the case of two-dimensional diffusion, we perturb the homogeneous steady state of the local system (\ref{eq:diff2}) around $(u^{*},v^{*})$ by 
\mathcenter
\begin{align*}
    \begin{pmatrix}
        u \\
        v
    \end{pmatrix}=\begin{pmatrix}
        u^{*} \\
        v^{*}
    \end{pmatrix}+\epsilon \begin{pmatrix}
        u_{1} \\
        v_{1}
    \end{pmatrix}\exp{(\lambda t+i(k_{x}x+k_{y}y))},
\end{align*}
where $|\epsilon|\ll 1$ and $\lambda$ is the growth rate of perturbation. 
Substituting the aforementioned transformations in \eqref{eq:diff2}, we get the following linearized form

\begin{equation} \label{eq:3.2}
\textbf{J}_{k}\begin{bmatrix}
u_{1}\\
v_{1}
\end{bmatrix}\equiv \begin{bmatrix}
 a_{11}-d_{1}k^{2}-\lambda & a_{12} \\
 a_{21} & a_{22}-d_{2}k^{2}-\lambda 
\end{bmatrix}
\begin{bmatrix}
    u_{1} \\
    v_{1}
\end{bmatrix}
=\begin{bmatrix}
        0 \\
        0
    \end{bmatrix},
\end{equation}
where $a_{11},\ a_{12},\ a_{21}$ and $a_{22}$ are provided in the subsection \ref{subsec-2.3}. And, $\textbf{k}=(k_{x},k_{y})$ is the wave number vector along with the wave number $k=|\textbf{k}|$. We are interested in finding the non-trivial solution of the system (\ref{eq:3.2}), so $\lambda$ must be a zero of $\det(\textbf{J}_{k})=0$, which gives 
\begin{align*}
    \lambda_{\pm}(k^{2})=\frac{\Phi_{1}(k^{2})\pm\sqrt{(\Phi_{1}(k^{2}))^{2}-4\Phi_{2}(k^{2})}}{2},
\end{align*}
where $\Phi_{1}(k^{2})=\mbox{tr}(\textbf{J}(E^{*}))-(d_{1}+d_{2})k^{2},\ \Phi_{2}(k^{2})=\det(\textbf{J}(E^*))-(d_{2}a_{11}+d_{1}a_{22})k^{2}+d_{1}d_{2}k^{4}$.
In the absence of species diffusion, the coexisting equilibrium is locally stable for $\mbox{tr}(\textbf{J}(E^{*}))<0$ and $\det(\textbf{J}(E^*))>0$. It indicates that $\Phi_{1}(k^{2})<0$ for all $k$ when the temporal model is locally asymptotically stable. So, the homogeneous solution will be stable under heterogeneous perturbation when $\Phi_{2}(k^{2})>0$ for all $k$. If the inequality is violated for some $k \neq 0$, the system is unstable.

Here, $\displaystyle k^{2}_{min}=(d_{2}a_{11}+d_{1}a_{22})/2d_{1}d_{2}$ is the minimum value of $k^{2}$ for which $\displaystyle \Phi_{2}(k^{2})$ attains its minimum value as
$$\displaystyle \Phi_{2}(k^{2})_{min}=(a_{11}a_{22}-a_{12}a_{21})-\frac{(d_{2}a_{11}+d_{1}a_{22})^{2}}{4d_{1}d_{2}}.$$ 
This $k_{min}$ is the critical wave number for Turing instability. And the critical diffusion coefficient (Turing bifurcation threshold) $d_{1c}$ such that $\Phi_{2}(k^{2})_{min}=0$ is given as 
\begin{equation} \label{eq:3.5}
d_{1c}=\frac{d_{2}(a_{11}a_{22}-2a_{12}a_{21})-\sqrt{d_{2}^{2}(a_{11}a_{22}-2a_{12}a_{21})^{2}-d_{2}^{2}a_{11}^{2}a_{22}^{2}}}{a_{22}^{2}}.  
\end{equation}

For the asymptotic stability of the temporal system, the condition 
$a_{11}+a_{22}<0$ must hold. Now positivity at $k^{2}=k^{2}_{\min}$ requires that at least one of $a_{11}$ or $a_{22}$ be positive. Since $a_{22}<0$, it follows that $d_{1}<d_{2}$ must be satisfied to meet the conditions for Turing instability. This implies that the self-diffusion coefficient of the prey population is smaller than that of the predator population in the model (\ref{eq:diff2}). Furthermore, the positivity of $\Phi_{2}(k^{2})_{min}>0$ is achieved when $d_{1}<d_{1c}$, which facilitates the emergence of stationary and non-stationary patterns in the system. This indicates that the coexisting homogeneous steady-state $(u^{*},v^{*})$ of the local model (\ref{eq:diff2}) remains stable under random heterogeneous perturbations when $d_{1}>d_{1c}$. Consequently, the conditions for Turing instability are as follows:
\begin{equation}\label{eq:3.4}
(i) \ a_{11}+a_{22}<0, \ \ (ii) \ a_{11}a_{22}>a_{12}a_{21}, \ \mbox{and}\ (iii) \ d_{2}a_{11}+d_{1}a_{22}>2\sqrt{d_{1}d_{2}(a_{11}a_{22}-a_{12}a_{21})}. 
\end{equation}


\section{Spatial model and its behaviour}\label{sec:4}
Apart from the spatially uniform steady states, the
model \eqref{eq:diff2} can also admit spatially heterogeneous
steady states of the form $(u(\mathbf{x}),v(\mathbf{x}))$ which satisfy the following system:
\begin{equation} \label{eq:21}
\begin{aligned}
    -d_{1}\Delta u &=u(1-u)-\frac{f(v)uv}{m+hf(v)u+\gamma v} \equiv \overline{F}_{1}(u,v), \ \ \mathbf{x}\in\Omega, \\
    -d_{2}\Delta v &=\frac{\sigma f(v)uv}{m+hf(v)u+\gamma v}-\delta v \equiv \overline{F}_{2}(u,v), \ \ \mathbf{x}\in\Omega,
\end{aligned}
\end{equation}
subject to the no-flux boundary conditions. In this section, we establish results regarding the non-existence and existence of these spatially heterogeneous steady states. Now, system \eqref{eq:21} can be written as follows:
\begin{equation*}
\begin{aligned}
&-\Delta \mathbf{w}=\mathbf{F}(\mathbf{w}),\ \ x\in \Omega, \\
&\frac{\partial \mathbf{w}}{\partial\nu} = 0,\ \  x\in\partial\Omega,
\end{aligned}
\end{equation*}
where $\mathbf{w}=(u,v)$ and $\mathbf{F}(\mathbf{w})=(d_{1}^{-1}\overline{F}_{1}(\mathbf{w}), \ d_{2}^{-1}\overline{F}_{2}(\mathbf{w}))^{T}$. We consider $0=\lambda_{0}<\lambda_{1}<\lambda_{2}<\cdots<\lambda_{i}<\cdots$ as the eigenvalues of the operator $-\Delta$ on $\Omega$ with no-flux boundary conditions and $\varepsilon(\lambda_{i})$ be the eigenfunction space corresponding to the eigenvalue $\lambda_{i}$ in $C^{1}(\overline{\Omega})$. Let, $\{\phi_{ij}: j= 1,2,\ldots,\dim(\varepsilon(\lambda_{i}))\}$ be an orthonormal basis of $\varepsilon(\lambda_{i}),\ X= \{w\in C^{1}(\overline{\Omega})\times C^{1}(\overline{\Omega}): \partial w/\partial\nu=0\}$ on $\partial\Omega$ and $Xij=\{c\phi_{ij}: c\in\mathbb{R}^{2}\}$. Then, we have
$$X=\bigoplus_{i=1}^{\infty} X_{i},\ \mbox{where}\ X_{i}=\bigoplus_{j=1}^{\varepsilon(\lambda_{i})} X_{ij}.$$

\subsubsection{A priori estimate of positive steady state}

Now, we discuss the conditions regarding the non-existence of spatially heterogeneous steady states. The following two lemmas can be found in \cite{lin1988large, lou1999diffusion}, respectively. For this purpose, first, we deduce a priori estimates for non-negative solutions of the system \eqref{eq:21}.




\begin{lemma} \label{lem-1}
Let us assume that $(u(\mathbf{x}), v(\mathbf{x}))$ be a
non-negative solution of the system \eqref{eq:21}. Then, the solution $(u(\mathbf{x}), v(\mathbf{x}))$ satisfies $0<u(\mathbf{x})<u_{1}\equiv 1$ and $0<v(\mathbf{x})<v_{1}\equiv d_{1}\sigma+d_{2}\frac{\sigma}{\delta}$.
\end{lemma}

\begin{proof}
 If there exists $\mathbf{x}_{0}\in\Omega$ such that $u(\mathbf{x}_{0})=0$, then the strong maximum principle implies $u(\mathbf{x})=0$. Similarly, we have $v(\mathbf{x})=0$ when $v(\mathbf{x}_{0})=0$ for some $\mathbf{x}_{0}\in\Omega$. Otherwise, we have $u(\mathbf{x})>0$ and $v(\mathbf{x})>0$ for $\mathbf{x}\in\Omega$. 

Further from Theorem \ref{theorem-3.3}, we obtain $u(\mathbf{x})\leq 1$ as $t\rightarrow\infty$ for all $\mathbf{x}\in\Omega$. Then, using the strong maximum principle, we get $u(\mathbf{x})< 1$ for all $\mathbf{x}\in\Omega$. Now from \eqref{eq:21}, we obtain

\begin{align*}
    -(\sigma d_{1}\Delta u+d_{2}\Delta v)= \sigma u(1-u)-\delta v &=-\frac{\delta}{d_{2}}(\sigma d_{1}u+d_{2}v)+\sigma u\left[(1-u)+\frac{\delta d_{1}}{d_{2}}\right] \\
    &\leq -\frac{\delta}{d_{2}}(\sigma d_{1}u+d_{2}v)+\sigma u_{1}\left[1+\frac{\delta d_{1}}{d_{2}}\right]
\end{align*}
Then, using the maximum principle, we obtain
\begin{align*}
    \sigma d_{1}u+d_{2}v < \frac{d_{2}\sigma}{\delta}\left[1+\frac{\delta d_{1}}{d_{2}}\right]=d_{1}\sigma+d_{2}\frac{\sigma}{\delta}.
\end{align*}
Thus we get the estimate of $v(\mathbf{x})$ as $0<v(\mathbf{x})<v_{1}\equiv d_{1}\sigma+d_{2}\sigma/\delta$. Hence, proved.
\end{proof}

\subsection{Non-existence of non-constant positive steady-states}
Now, we present the result, which deals with the non-existence of heterogeneous steady-state solutions to the problem (\ref{eq:diff2}) when the diffusion coefficient $d_{1}$ varies while the other parameters $d_{2},\ \Gamma$ are fixed, where $\Gamma=(\omega, \phi, m, h, \gamma, \sigma, \delta, |\Omega|)$.

\begin{theorem}\label{Theorem-5}
System \eqref{eq:21} does not possess any non-constant positive solution for $d^{*}<\min\{d_{1},d_{2}\}$, where $d^{*}$ depends on parameter set $\Gamma$.
\end{theorem}

\begin{proof}
Let $(u(\mathbf{x}), v(\mathbf{x}))$ be a non-negative solution
of the system \eqref{eq:21}. Also, let us consider $\widehat{u}=\frac{1}{|\Omega|}\int_{\Omega}u(\mathbf{x})d\mathbf{x}$ and $\widehat{v}=\frac{1}{|\Omega|}\int_{\Omega}v(\mathbf{x})d\mathbf{x}$. Then, we have $$\int_{\Omega}(u(\mathbf{x})-\widehat{u})d\mathbf{x}=0,\ \ \int_{\Omega}(v(\mathbf{x})-\widehat{v})d\mathbf{x}=0$$
for any $u, v \in L^{1}(\Omega)$. Multiplying the first equation of \eqref{eq:21} by $(u(\mathbf{x})-\widehat{u})$ and integrating it over $\Omega$ along with Green’s identity and no-flux boundary condition, we obtain
\begin{equation*}
\begin{aligned}
d_{1}&\int_{\Omega}|\nabla(u-\widehat{u})|^{2}d\mathbf{x}\\
&=\int_{\Omega}\overline{F}_{1}(u,v)(u-\widehat{u})d\mathbf{x} \\
&=\int_{\Omega}(u-\widehat{u})\left[u(1-u)-\widehat{u}(1-\widehat{u})-\frac{f(v)uv}{m+hf(v)u+\gamma v}+\frac{f(\widehat{v})\widehat{u}\widehat{v}}{m+hf(\widehat{v})\widehat{u}+\gamma\widehat{v}}\right]d\mathbf{x} \\
&\leq \int_{\Omega}(u-\widehat{u})^{2}d\mathbf{x}+\bigintsss_{\Omega}\left[\frac{-m\widehat{u}(1-\omega)-hu\widehat{u}f(v)f(\widehat{v})+\frac{\gamma\omega\phi\widehat{v}uv}{(\phi+v)(\phi+\widehat{v})}}{(m+hf(v)u+\gamma v)(m+hf(\widehat{v})\widehat{u}+\gamma\widehat{v})}\right](u-\widehat{u})(v-\widehat{v})d\mathbf{x} \\
&\leq \int_{\Omega}(u-\widehat{u})^{2}d\mathbf{x}+\frac{1}{m^{2}}[m(1-\omega)\widehat{u}+\gamma\omega\phi+h\widehat{u}]\int_{\Omega}|(u-\widehat{u})||(v-\widehat{v})|d\mathbf{x} 
\end{aligned}
\end{equation*}
From Lemma \ref{lem-1}, we have $\displaystyle \int_{\overline{\Omega}}u(\mathbf{x})d\mathbf{x} \leq |\Omega|$. This gives $\widehat{u}=\frac{1}{|\Omega|}\int_{\Omega}u(\mathbf{x})d\mathbf{x}\leq 1$. Choose $\mu_{1}=m(1-\omega)+\gamma\omega\phi+h$. Hence, 
\begin{equation}\label{eq:22}
\begin{aligned}
d_{1}\int_{\Omega}|\nabla(u-\widehat{u})|^{2}d\mathbf{x}&\leq \int_{\Omega}(u-\widehat{u})^{2}d\mathbf{x}+\frac{\mu_{1}}{m^{2}}\int_{\Omega}|u-\widehat{u}||v-\widehat{v}|d\mathbf{x} \\
\end{aligned}
\end{equation}
Again, multiplying the second equation of \eqref{eq:21} by $(v(\mathbf{x})-\widehat{v})$ and integrating it over $\Omega$ along with no-flux boundary condition, we obtain
\begin{equation*}
\begin{aligned}
d_{2}\int_{\Omega}|\nabla(v-\widehat{v})|^{2}d\mathbf{x}&=\int_{\Omega}v(v-\widehat{v})\left[\frac{\sigma f(v)u}{m+hf(v)u+\gamma v}-\delta \right]d\mathbf{x} \\
&=\int_{\Omega}[(v-\widehat{v})^{2}+\widehat{v}(v-\widehat{v})]\left[\frac{\sigma f(v)u}{m+hf(v)u+\gamma v}-\delta \right]d\mathbf{x}
\end{aligned}
\end{equation*}
\mathleft 
\begin{align*}
\textrm{From \eqref{eq:21} we obtain}\   \int_{\Omega}-(\sigma d_{1}\Delta u+d_{2}\Delta v)d\mathbf{x}=\int_{\Omega}[\sigma u(1-u)-\delta v]d\mathbf{x}.
\end{align*}
\mathcenter
Using the no-flux boundary conditions we get
\begin{align*}
\delta\int_{\Omega}vd\mathbf{x}=\int_{\Omega}[\sigma u(1-u)]d\mathbf{x} \leq \frac{\sigma}{4}|\Omega|.
\end{align*}

which gives $\displaystyle \widehat{v}=\frac{1}{|\Omega|}\int_{\Omega}vd\mathbf{x}\leq \frac{\sigma}{4\delta}$.
Hence, we get 
\begin{equation}\label{eq:23}
\begin{aligned}
d_{2}\int_{\Omega}|\nabla(v-\widehat{v})|^{2}d\mathbf{x}&\leq \left(\frac{\sigma}{h}-\delta\right)\int_{\Omega}(v-\widehat{v})^{2}d\mathbf{x}+\frac{\sigma^{2}}{4\delta m^{2}}\left(1+\frac{\sigma}{4\delta}\right)\int_{\Omega}|u-\widehat{u}||v-\widehat{v}|d\mathbf{x} \\
\end{aligned}
\end{equation}
From \eqref{eq:22} and \eqref{eq:23}, we obtain 
\begin{align*}
d_{1}\int_{\Omega}|\nabla(u-\widehat{u})|^{2}d\mathbf{x}+d_{2}\int_{\Omega}|\nabla(v-\widehat{v})|^{2}d\mathbf{x} &\leq \int_{\Omega}\left(1+\mu_{3}\right)(u-\widehat{u})^{2}d\mathbf{x} \\
&~~~~+\int_{\Omega}\left(\frac{\sigma}{h}-\delta+\mu_{3}\right)(v-\widehat{v})^{2}d\mathbf{x}
\end{align*}
where $\mu_{3} = \frac{\mu_{1}}{2m^{2}}+\frac{\sigma^{2}}{8\delta m^{2}}\left(1+\frac{\sigma}{4\delta}\right)$. Then applying the Poincar$\acute{e}$ inequality, we obtain


\begin{equation}{\label{NOE}}
\int_{\Omega}\lambda_{1}[d_{1}(u-\widehat{u})^{2}+d_{2}(v-\widehat{v})^{2}]d\mathbf{x} \leq \int_{\Omega}\left[\left(1+\mu_{3}\right)(u-\widehat{u})^{2}+ \left(\frac{\sigma}{h}-\delta+\mu_{3}\right)(v-\widehat{v})^{2}\right]d\mathbf{x},
\end{equation}
where $\lambda_{1}$ is the smallest positive eigenvalue of the operator $-\Delta$ on $\Omega$ with no-flux boundary condition. Now, if $\min \{ d_{1}, d_{2}\} > \max\{ (1+\mu_{3})/\lambda_{1}, (\sigma/h-\delta+\mu_{3})/\lambda_{1} \} \equiv d^{*}$, then the inequality (\ref{NOE}) holds only for $u=\widehat{u}$ and $v=\widehat{v}$. Hence, proved.


\end{proof}

\subsection{Existence of positive non-constant steady states}
Here, we discuss the existence of spatially heterogeneous steady states (non-constant positive classical solutions) to (\ref{eq:21}) when the diffusion coefficient $d_{2}$ vary while keeping the fixed parameters $d_{1}$ and $\Gamma$. Theorem \ref{Theorem-3.5} implies that when the steady-state solution of \eqref{eq:det1} exists satisfies the conditions $a_{11}+a_{22}<0$ and $d_{1}a_{22}+d_{2}a_{11}<2\sqrt{d_{1}d_{2}(a_{11}a_{12}-a_{12}a_{21})}$, then (\ref{eq:21}) has no non-constant positive classical solutions. Given this reason, we shall restrict this discussion to the case where a positive steady-state solution exists and $d_{1}a_{22}+d_{2}a_{11}> 2\sqrt{d_{1}d_{2}(a_{11}a_{12}-a_{12}a_{21})}$.

In that account, we denote $\mathbf{w}=(u, v),\ \mathbf{w}^{*} = E^{*}=(u^{*}, v^{*})$ and $X^{+}=\{\mathbf{w}\in X : u,v> 0\ \textrm{on}\ \overline{\Omega}\}$. Then, the system \eqref{eq:21} becomes
\begin{equation}\label{eq:24}
\begin{aligned}
    -\Delta \mathbf{w}& = \mathbf{F}(\mathbf{w}),\ \mathbf{w}\in X^{+},\ \mathbf{x}\in \Omega, \\
    \frac{\partial \mathbf{w}}{\partial\nu}& = 0,~~   \mathbf{x}\in\partial\Omega,
\end{aligned}
\end{equation}
where $\mathbf{F}(\mathbf{w})=(d_{1}^{-1}\overline{F}_{1}(\mathbf{w}), \ d_{2}^{-1}\overline{F}_{2}(\mathbf{w}))^{T}$. Now, we define an operator $\mathcal{F}:X^{+}\rightarrow X^{+}$ as $$\mathcal{F}(\mathbf{w})=(I-\Delta)^{-1}\{\mathbf{F}(\mathbf{w})+\mathbf{w}\},$$
where $(I-\Delta)^{-1}$ denotes the inverse operator of $(I-\Delta)$ under the no-flux boundary condition. Then, the system \eqref{eq:24} possesses a positive solution $\mathbf{w}=(u,v)$ if and only if $\mathbf{w}$ is a positive solution to the equation
$$\mathcal{G}(d_{1},d_{2},\mathbf{w}) \equiv \mathbf{w}-\mathcal{F}(\mathbf{w}) = 0~~\mbox{for}~\mathbf{w}\in X^{+}.$$
Now, $$\mathcal{G}_{\mathbf{w}}(d_{1},d_{2},\mathbf{w}) = I-(I-\Delta)^{-1}\{\mathbf{F}_{\mathbf{w}}(\mathbf{w})+I\},$$
and if $\mathcal{G}_{\mathbf{w}}(d_{1},d_{2},\mathbf{w})$ is invertible, then, according to the Leray-Schauder method, the index of $\mathcal{G}$ at $\mathbf{w}^{*}$ is defined as:
$$\mbox{Index}~(\mathcal{G}(d_{1},d_{2},\mathbf{w}^{*})) = (-1)^{\rho},$$
where $\rho$ is the sum of the algebraic multiplicity of the negative eigenvalue of $\mathcal{G}_{\mathbf{w}}(d_{1},d_{2},\mathbf{w}^{*})$. If $\psi$ is an eigenvalue of $\mathcal{G}_{\mathbf{w}}(d_{1},d_{2},\mathbf{w}^{*})$ on $\varepsilon(\lambda_{i})$ if and only if $\psi(1+\lambda_{i})$ is an eigenvalue of the matrix:
\begin{align*}
    \lambda_{i}I-\mathcal{J}=\begin{pmatrix}
        \lambda_{i}-a_{11}/d_{1} & -a_{12}/d_{1} \\
        -a_{21}/d_{2} & \lambda_{i}-a_{22}/d_{2}
    \end{pmatrix}.
\end{align*} 
The invertibility of $\mathcal{G}_{\mathbf{w}}(d_{1},d_{2},\mathbf{w}^{*})$ depends on the condition $\mathcal{H}(d_{1},d_{2}; \lambda_{i}) \neq 0$ for $i\geq 0$, where $$\mathcal{H}(d_{1},d_{2}; \lambda_{i}) = d_{1}d_{2}\det(\lambda_{i}-\mathcal{J}) = d_{1}d_{2}\lambda_{i}^2-(d_{1}a_{22}+d_{2}a_{11})\lambda_{i}+(a_{11}a_{22}-a_{12}a_{21}).$$
Here, the roots of $\mathcal{H}(d_{1},d_{2}; \lambda)=0$ are
\begin{align*}
    \lambda_{\pm}(k^{2})=\frac{(d_{1}a_{22}+d_{2}a_{11})\pm\sqrt{(d_{1}a_{22}+d_{2}a_{11})^{2}-4d_{1}d_{2}(a_{11}a_{22}-a_{12}a_{21})}}{2d_{1}d_{2}}
\end{align*}
and they are real when $(d_{1}a_{22}+d_{2}a_{11})>2\sqrt{d_{1}d_{2}(a_{11}a_{22}-a_{12}a_{21})}$. With all these, we define $\mathcal{B}(d_{1},d_{2})=\{\lambda: \lambda\geq 0, \lambda_{-}<\lambda<\lambda_{+}\},\ \Lambda=\{\lambda_{0},\lambda_{1},\lambda_{2}, \ldots\}$, and $\mathcal{P}(\lambda_{i})$ be the algebraic multiplicity of $\lambda_{i}$. 

\begin{lemma}\label{lem-2}
Suppose $\mathcal{H}(d_{1}, d_{2}; \lambda_{i})\neq 0$ for all
$\lambda_{i}\in\Lambda$. Then $\mbox{Index}(\mathcal{G}(d_{1},d_{2},\mathbf{w}^{*}))=(-1)^{\rho}$, where 
\begin{align*}
    \rho=\begin{cases}
      \Sigma_{\lambda_{i}\in\mathcal{B}\cap\Lambda}\mathcal{P}(\lambda_{i})\ &\textrm{if}\ \mathcal{B}\cap\Lambda\neq \emptyset \\
      ~~~~~~~0 \ &\textrm{if}\ \mathcal{B}\cap\Lambda = \emptyset.
    \end{cases}      
\end{align*}
In particular, $\rho=0$ when $\mathcal{H}(d_{1}, d_{2}; \lambda_{i})>0$ for all $\lambda_{i}\geq 0$.
\end{lemma}


\begin{theorem}\label{Theorem-6}
Suppose the condition $(d_{2}a_{11}+d_{1}a_{22})>2\sqrt{d_{1}d_{2}(a_{11}a_{22}-a_{12}a_{21})}$ holds for a unique spatially uniform steady state $\mathbf{w}^{*}$ and there exist some integers $m$ and $n$ with $0\leq m< n$ such that $\lambda_{-}\in(\lambda_{m},\lambda_{m+1}),\ \lambda_{+}\in(\lambda_{n},\lambda_{n+1})$ and $\Sigma^{n}
_{i=m+1}\mathcal{P}(\lambda_{i})$ is odd, then the system \eqref{eq:21} admits at least one positive spatially heterogeneous solution.
\end{theorem}

\begin{proof}
Using the Lemma \ref{lem-1}, there exists a real number $M>0$ such that $M^{-1}<u, v<M$ for $\mathbf{x}\in\Omega$ and we define $$\mathbb{B}=\{(u,v)\in X: M^{-1}<u,\ v<M\ \textrm{for } \mathbf{x}\in\Omega\}.$$


Theorem \ref{Theorem-5} shows that the system \eqref{eq:21} has no non-constant positive solution for some $d^{*}<\min\{d_{1},d_{2}\}$. Suppose, on the contrary, the system \eqref{eq:21} has no non-constant positive solution even for some $d^{*}>\min\{d_{1},d_{2}\}$. Now, we consider $\widehat{\mathcal{F}}: \mathbb{B}\times[0,1]\rightarrow C(\overline{\Omega})\times C(\overline{\Omega})$ given by
$\widehat{\mathcal{F}}(\mathbf{w}, \tau):=(I-\Delta)^{-1}\{\mathbf{G}(\mathbf{w},\tau)+\mathbf{w}\}$, where 
\begin{align*}
\mathbf{G}(\mathbf{w},\tau)=\begin{bmatrix}
    (\tau d_{1}+(1-\tau)d^{*})^{-1}\overline{F}_{1}(u,v) \\
    (\tau d_{2}+(1-\tau)d^{*})^{-1}\overline{F}_{2}(u,v)
\end{bmatrix}    
\end{align*}
with $\tau\in[0,1]$. This shows that finding a positive spatially heterogeneous
solution of the system \eqref{eq:24} becomes the same as finding a fixed point of $\widehat{\mathcal{F}}(\mathbf{w},1)$ in $\mathbb{B}$. 
In addition, the operator $\widehat{\mathcal{F}}(\mathbf{w},\tau)$ does not possess a fixed point on $\partial\mathbb{B}$ for $\tau\in[0, 1]$. Furthermore, the Leray-Schauder degree $\deg(I-\widehat{\mathcal{F}}(\mathbf{w},\tau),\mathbb{B},0)$ is well-defined for $\tau\in[0, 1]$ as $\widehat{\mathcal{F}}(\mathbf{w},\tau)$ is compact. Therefore, the homotopy invariance of the Leray-Schauder degree gives 
\begin{equation}\label{eq:26}
\deg(I-\widehat{\mathcal{F}}(\mathbf{w},1),\mathbb{B},0)=\deg(I-\widehat{\mathcal{F}}(\mathbf{w},0),\mathbb{B},0)   
\end{equation}
when the system possesses a unique $\mathbf{w}^{*}$ in $\mathbb{B}$.\\
If the condition $(d_{2}a_{11}+d_{1}a_{22})>2\sqrt{d_{1}d_{2}(a_{11}a_{22}-a_{12}a_{21})}$ holds for a unique spatially uniform steady state $\mathbf{w}^{*}$ of the system \eqref{eq:21}, then we obtain $$\mathcal{B}(d_{1}, d_{2})\cap\Lambda= \{\lambda_{m+1}, \lambda_{m+2},\ldots,\lambda_{n}\}~~\mbox{and}~~\mathcal{B}(d^{*}, d^{*})\cap\Lambda=\emptyset$$ since $\lambda_{-}\in(\lambda_{m}, \lambda_{m+1})$ and $\lambda_{+}\in(\lambda_{n},\lambda_{n+1})$.

Therefore, at $\tau = 0$, by applying Lemma \ref{lem-2}, we get:
\begin{equation*}\label{eq:27}
\deg(I-\widehat{\mathcal{F}}(\mathbf{w},0),\mathbb{B},0)= \mbox{Index}(I-\widehat{\mathcal{F}}(\mathbf{w},0),\mathbf{w}^{*}) =(-1)^{0}=+1.
\end{equation*}

On the other hand, $\tau = 1$ gives $I-\widehat{\mathcal{F}}(\mathbf{w},1)=I-\mathcal{F}(\mathbf{w})$ and if $\Sigma^{n}_{i=m+1}\mathcal{P}(\lambda_{i})$ is odd, then again from Lemma \ref{lem-2}, we deduce 
\begin{equation*}\label{eq:28}
\deg(I-\widehat{\mathcal{F}}(\mathbf{w},1),\mathbb{B},0)= \mbox{Index}(I-\widehat{\mathcal{F}}(\mathbf{w},1),\mathbf{w}^{*}) =(-1)^{\Sigma^{n}_{i=m+1}\mathcal{P}(\lambda_{i})}=-1
\end{equation*}
if the system \eqref{eq:24} admits only the positive constant solution $\mathbf{w}^{*}$. Thus, we contradict the degree-invariance  \eqref{eq:26}. Hence, the system \eqref{eq:21} admits at least one positive spatially heterogeneous solution for $\min\{d_{1}, d_{2}\}<d^{*}$.
\end{proof}



\section{Analysis of taxis-driven instability} \label{sec:5}

\begin{theorem}
Suppose the temporal model \eqref{eq:det1} possesses a feasible interior equilibrium point $E^{*}$ which is locally asymptotically stable. Then, the condition 
\begin{equation}\label{eq:4.3}
 d_{2}a_{11}+d_{1}a_{22}+(a_{12}\eta v^{*}-a_{21}\xi u^{*})>0   
\end{equation}
needs to be held for Turing instability in system \eqref{eq:diff3}.
Moreover, if $d_{2}a_{11}+d_{1}a_{22}+a_{12}\eta v^{*}>2\sqrt{d_{1}d_{2}(a_{11}a_{22}-a_{12}a_{21})}$
holds, then there exists a critical threshold $\xi_{c}$ such that $E^{*}$ is locally asymptotically stable when $\xi>\xi_{c}$ and unstable when $0<\xi<\xi_{c}$. In addition, the system exhibits a taxis-driven Turing bifurcation at $\xi=\xi_{c}$ and a Turing pattern emerges in the system for $0<\xi<\xi_{c}$ when \eqref{eq:4.3} is satisfied.
\end{theorem}

\begin{proof}
We consider a heterogeneous perturbation around the stable coexisting homogeneous steady state $E^{*}$. After linearizing around $E^{*}$, the spatio-temporal system \eqref{eq:diff3} takes the following form:
\begin{equation} \label{eq:4.2}
\textbf{M}_{k}\begin{bmatrix}
u_{1}\\
v_{1}
\end{bmatrix}\equiv \begin{bmatrix}
 a_{11}-d_{1}k^{2}-\lambda & a_{12}-\xi u^{*}k^{2} \\
 a_{21}+\eta v^{*}k^{2} & a_{22}-d_{2}k^{2}-\lambda 
\end{bmatrix}
\begin{bmatrix}
    u_{1} \\
    v_{1}
\end{bmatrix}
=\begin{bmatrix}
        0 \\
        0
    \end{bmatrix},
\end{equation}
where 
$a_{11},\ a_{12},\ a_{21}$ and $a_{22}$ are mentioned in the subsection \ref{subsec-2.3}. To find the non-trivial solution of the system (\ref{eq:4.2}), $\lambda$ has to be a zero of $\det(\textbf{J}_{k})=0$, which reduces to
\begin{align*}
    \det(\textbf{M}_{k})=\lambda^{2}-\zeta_{1}(k^{2})\lambda+\zeta_{2}(k^{2}) = 0,
\end{align*}
where $\zeta_{1}(k^{2})=(a_{11}+a_{22})-(d_{1}+d_{2})k^{2} 
$ and $\zeta_{2}(k^{2})=(d_{1}d_{2}+\xi\eta u^{*}v^{*})k^{4}-[(d_{2}a_{11}+d_{1}a_{22})+(a_{12}\eta v^{*}-a_{21}\xi u^{*})]k^{2}+(a_{11}a_{22}-a_{12}a_{21})$. The term $\zeta_{1}(k^{2})>0$ always holds for all $k$ when \eqref{eq:2.6} is satisfied, and hence, the local asymptotic stability of $E^{*}$ for system \eqref{eq:diff3} is determined by the sign of $\zeta_{2}(k^{2})$.  Suppose, if $d_{2}a_{11}+d_{1}a_{22}+a_{12}\eta v^{*}\leq a_{21}\xi u^{*}$ holds, then $\zeta_{2}(k^{2})>0$ for all $k$ under the assumptions $a_{11}+a_{22}<0$ and $a_{11}a_{22}>a_{12}a_{21}$. In this case, the necessary condition for the existence of Turing patterns for positive $\xi$ and $\eta$ is given by:  
\begin{equation*}
d_{2}a_{11}+d_{1}a_{22}+(a_{12}\eta v^{*}-a_{21}\xi u^{*})>0. 
\end{equation*}
\mathcenter

Now, we find the critical taxis-driven Turing bifurcation threshold for which the minimum of $\zeta_{2}(k^{2})$ satisfies the condition $\min_{k^{2}>0}\zeta_{2}(k^{2})=0.$
Here, $\zeta_{2}$ attains its minimum value at the critical wave number $k^{2}=\overline{k}_{c}^{2}$, where
$$\overline{k}_{c}^{2}=\frac{(d_{2}a_{11}+d_{1}a_{22})+a_{12}\eta v^{*}-a_{21}\xi u^{*}}{2(d_{1}d_{2}+\xi\eta u^{*}v^{*})}>0.$$ 
Using condition \eqref{eq:4.3}, it is obtained that the critical predator-taxis is a solution of the following equation:
\begin{equation} \label{eq:4.4}
\begin{aligned}
a_{11}a_{22}-a_{12}a_{21}&=\frac{[(d_{2}a_{11}+d_{1}a_{22})+(a_{12}\eta v^{*}-a_{21}\xi u^{*})]^{2}}{4(d_{1}d_{2}+\xi\eta u^{*}v^{*})} 
\end{aligned}
\end{equation}

\begin{equation*} 
\begin{aligned}
\textrm{which gives}\ \ &\xi_{c}=\frac{\rho_{1}(\eta,d_{1},d_{2})- \sqrt{\{\rho_{1}(\eta,d_{1},d_{2})\}^{2}-(a_{21}u^{*})^{2}\rho_{2}(\eta,d_{1},d_{2})}}{(a_{21}u^{*})^{2}}
\end{aligned}
\end{equation*}
where $\rho_{1}(\eta,d_{1},d_{2})=a_{21}u^{*}(d_{2}a_{11}+d_{1}a_{22}+a_{12}\eta v^{*})+2\eta u^{*}v^{*}(a_{11}a_{22}-a_{12}a_{21})$ and $\rho_{2}(\eta,d_{1},d_{2})=(d_{2}a_{11}+d_{1}a_{22}+a_{12}\eta v^{*})^{2}-4d_{1}d_{2}(a_{11}a_{22}-a_{12}a_{21})$. 
Now, positivity of $\xi_{c}$ holds when $\rho_{2}(\eta,d_{1},d_{2})>0$, and it indicates $d_{2}a_{11}+d_{1}a_{22}+a_{21}\eta v^{*}>2\sqrt{d_{1}d_{2}(a_{11}a_{22}-a_{12}a_{21})}$. Moreover, $\min_{k^{2}>0}\zeta_{2}(k^{2})<0$ and \eqref{eq:4.3} hold simultaneously whenever $\xi<\xi_{c}$, which leads to Turing instability. On the other hand, when $\xi>\xi_{c}$, we have $\min_{k^{2}>0}\zeta_{2}(k^{2})>0$. It should be noted that $\zeta_{1}(k^{2})>0$ for all $k>0$ when the conditions in \eqref{eq:2.6} hold. Henceforth, system \eqref{eq:diff3} is locally asymptotically stable for $\xi>\xi_{c}$.


\end{proof}

\section{Numerical Results} \label{sec:6}

\begin{figure}[htb!]
    \centering
    \includegraphics[width=10cm]{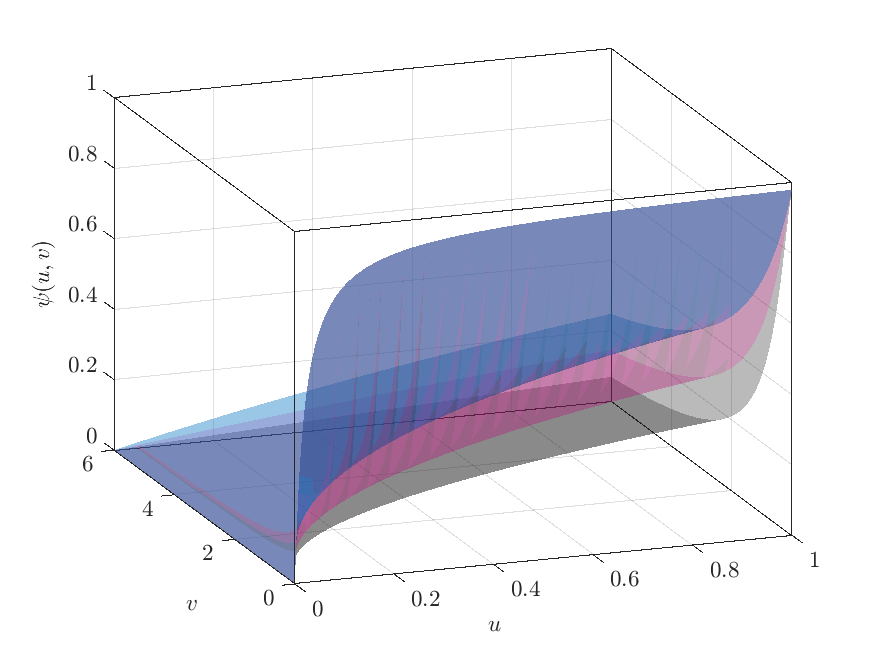}
    \caption{The effect of inducible defence on Beddington-DeAngelis functional response $\psi(u,v)=uf(v)/(m+f(v)u+\gamma v)$ with $f(v)=1-\omega v/(\phi+v)$. The surfaces plotted in (\legendsquare{fadblu1}), (\legendsquare{mag1}), and (\legendsquare{gray1}) colors represent the consumption rate function for $\omega=0, 0.5$, and $0.8$, respectively. Fixed parameter values are given in Table \ref{Table:1}.} \label{fig:1}
\end{figure}

This section contains the numerical validation of the dynamic behaviour of the proposed predator-prey model in the presence and absence of taxis. In the predator-prey interaction, the prey species exhibits an inducible defence strategy against their predators. Our goal is to find out the prominence or influence of this type of defence in the presence of predator interference. Figure \ref{fig:1} shows the effect of the defence mechanism on the Beddington-DeAngelis functional response. To perform the simulation, we have fixed some of the parameters used in the model, which are given in Table \ref{Table:1}. It is observed in Fig. \ref{fig:1} that the consumption rate plummets in the presence of defence, and hence, increasing prey biomass makes more impact on the consumption rate than their predators. 

\begin{table}[!ht]
\centering
\begin{tabular}{|c|c|c|c|c|c|}
\hline
Parameters & $\phi$ & $m$ & $h$ & $\sigma$ & $\delta$ \\
 \hline
Value & $0.2$ & $0.02$ & $1$ & $0.95$ & $0.43$ \\
\hline
\end{tabular}
\caption{Fixed parameter values used in the numerical simulations.} \label{Table:1}
\end{table}

\begin{figure}[htb!]
    \centering
     \centering
    \begin{subfigure}[t]{0.5\textwidth}
        \centering
        \includegraphics[width=\linewidth]{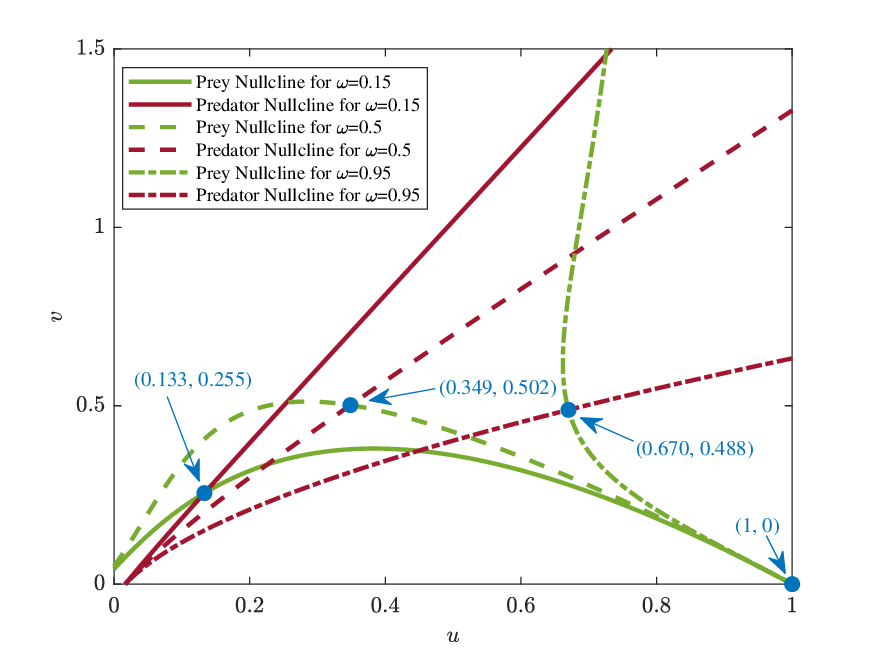}
        \caption{}\label{fig:2a}
    \end{subfigure}%
    ~ 
    \begin{subfigure}[t]{0.5\textwidth}
        \centering
        \includegraphics[width=\linewidth]{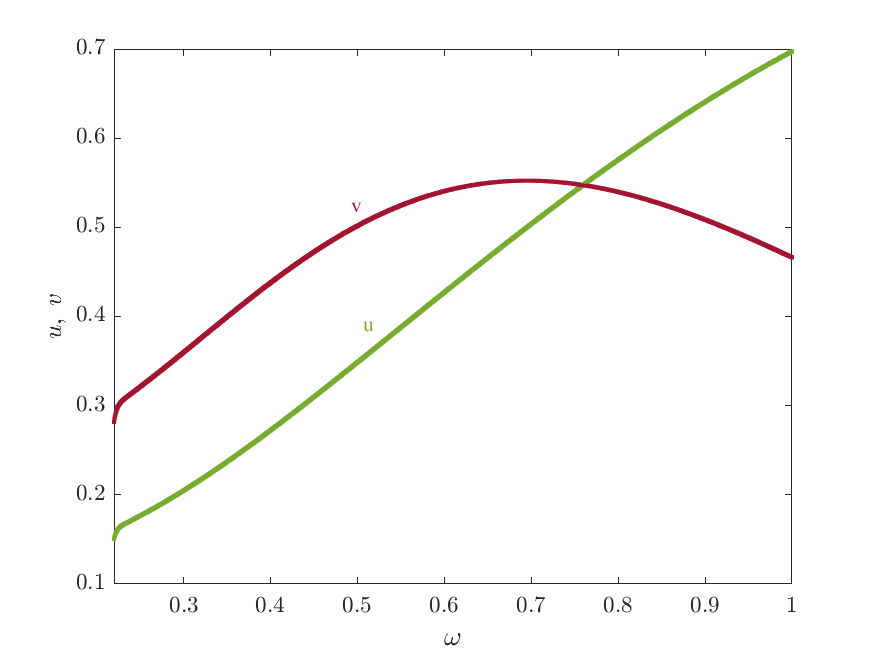}
        \caption{}\label{fig:2b}
    \end{subfigure}
    \caption{(a) The non-trivial prey are predator nullclines are represented when $\omega=0$ and $\omega\neq 0$. (b) The non-trivial equilibrium component of prey $(u)$ and predator $(v)$ species for increasing inducible defence. Fixed parameter values are given in Table \ref{Table:1}.} \label{fig:2}
\end{figure}

Figure \ref{fig:2} shows that the prey density gets immoderately affected when they show inducible defence. The biomass gets higher at a striking rate for increasing $\omega$. On the other hand, the predator goes up even in the presence of a lower defence level, but if the prey starts to show a higher defence, the predator biomass ultimately decreases. 

\subsection{Temporal bifurcations}

\begin{figure}[htb!]
    \centering
    \begin{subfigure}[t]{.45\textwidth}
        \centering
        \includegraphics[width=\linewidth]{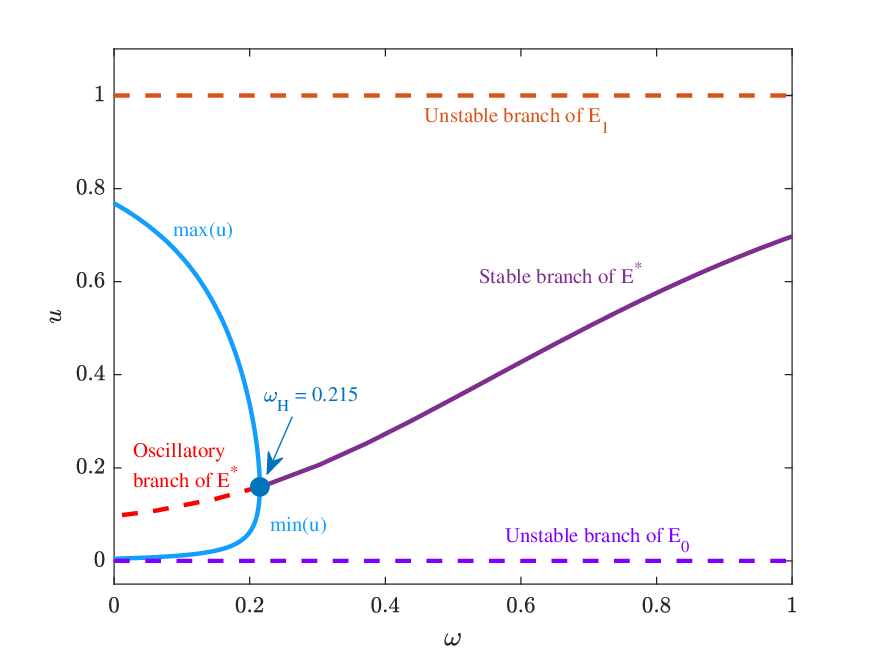}
        \caption{}\label{fig:3a}
    \end{subfigure}
    \begin{subfigure}[t]{.45\textwidth}
        \centering
        \includegraphics[width=\linewidth]{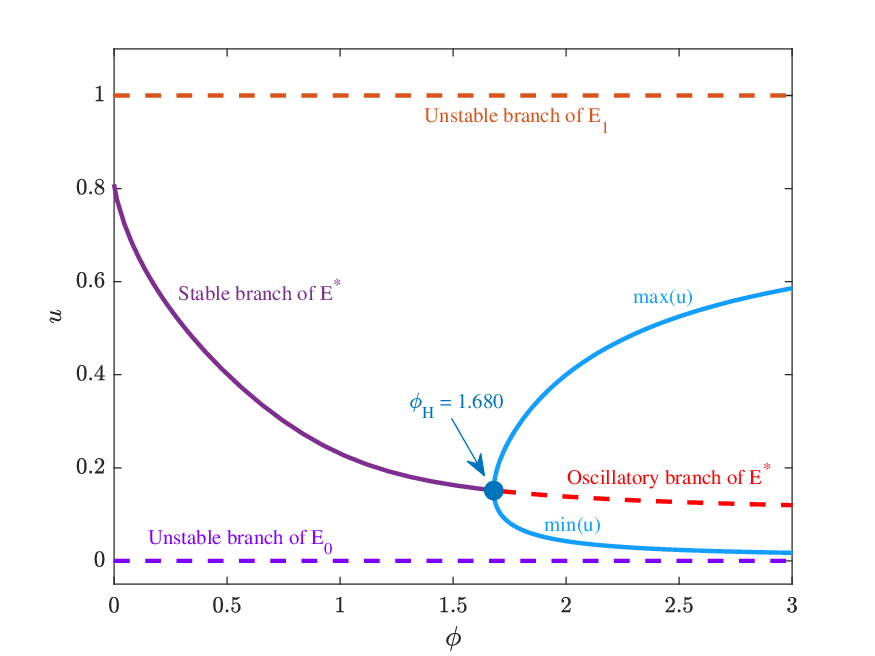}
        \caption{}\label{fig:3b}
    \end{subfigure}
    \begin{subfigure}{.45\textwidth}
        \centering
        \includegraphics[width=\linewidth]{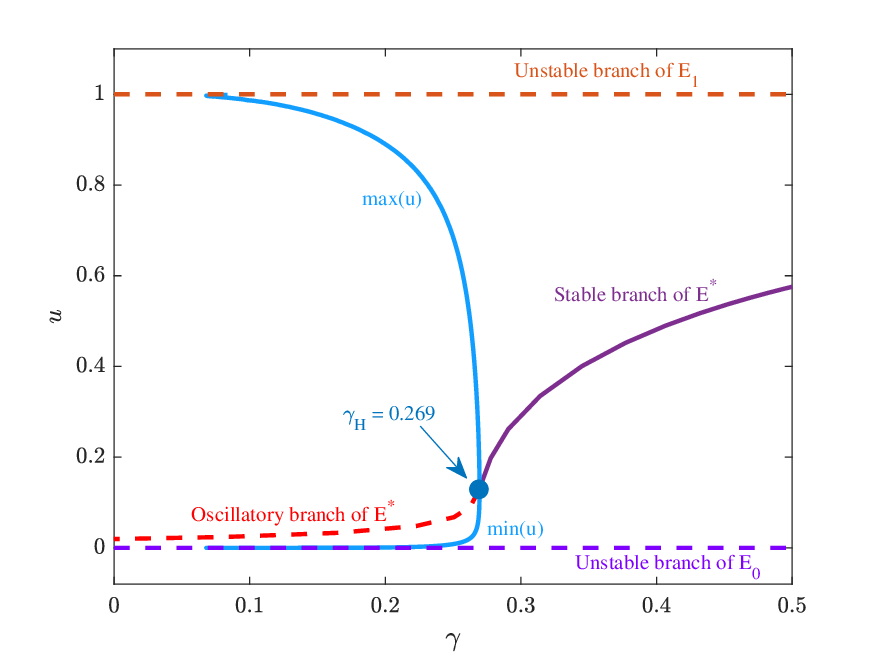}
        \caption{}\label{fig:3c}
    \end{subfigure}
    \begin{subfigure}{.45\textwidth}
        \centering
        \includegraphics[width=\linewidth]{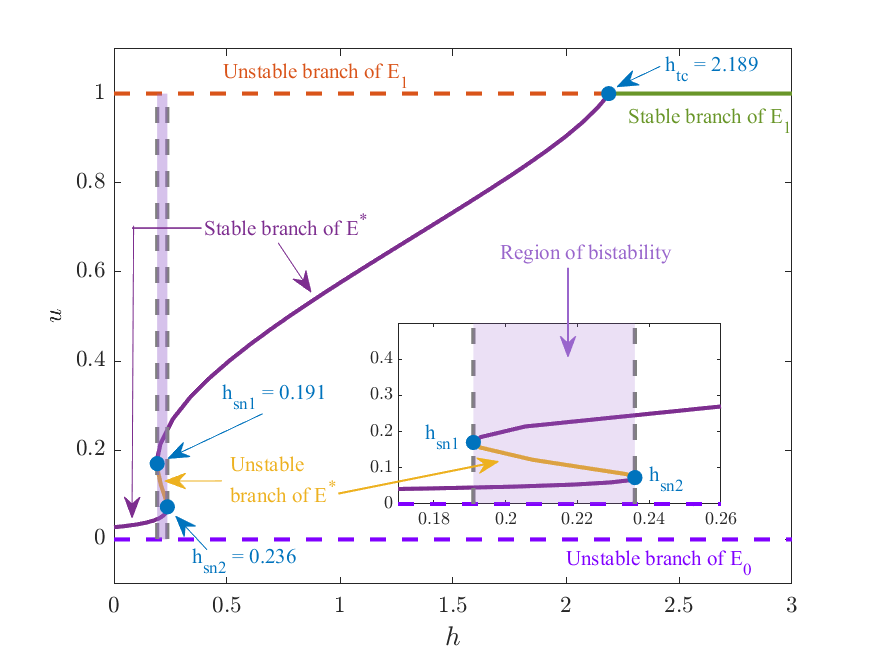}
        \caption{}\label{fig:3d}
    \end{subfigure}
    \caption{Bifurcations of the equilibrium points of the temporal model (\ref{eq:det1}) choosing (a) $\omega$, (b) $\phi$, (c) $\gamma$, and (d) $h$ as the bifurcation parameters. } \label{fig:3}
\end{figure}

Before analyzing the spatio-temporal behaviours of the model, it is important to see how the system parameters affect the dynamic scenario of the temporal system. The bifurcation diagram in Fig. \ref{fig:3a} shows that the inducible defence of prey species mainly acts as a stabilizing factor in the system as the species, from oscillation, settles down to a stable coexisting state for increasing $\omega$, and the system undergoes a supercritical Hopf bifurcation around $E^{*}$ at $\omega_{H}=0.215$. So, it illustrates that for a lower defence level, the species show oscillations, indicating regular fluctuations. However, the prey population can endure the predator pressure more effectively due to the stronger defence, leading to reduced oscillation and ultimately reaching stable coexistence with predators. This stability switching actually inhibits the higher oscillations, stabilizes the population, and leads to a more resilient ecosystem.  

Next, we choose the parameter $\phi$ as a regulating parameter. Figure \ref{fig:3b} shows that the population acts as a stable interior point for a minimal value of $\phi$ but starts to oscillate while crossing a threshold value. This stability switching occurs through a supercritical Hopf bifurcation around $E^{*}$ at $\phi_{H}=1.680$. On the other hand, Fig. \ref{fig:3c} portrays how predator interference makes an impact on the dynamic nature of the system. It is observed that the species show oscillation when the predators interfere at a lower rate. However, increasing interference lowers the oscillation amplitude, and the system tends to a stable coexisting state. In this case, the system undergoes a supercritical Hopf bifurcation around $E^{*}$ at $\gamma_{H}=0.269$. This result indicates that the species fluctuate over time when there is a minimal rate of predator interference present, representing a balanced ecosystem, but while the interference occurs at a higher rate, this oscillation settles to such an equilibrium when both prey and predator stabilize at constant levels. Figure \ref{fig:3d} depicts interesting dynamics where the prey handling time $(h)$ is chosen as the regulating parameter. It is observed that the predator-free state $(E_{1})$ changes its stability from unstable to stable state through a transcritical bifurcation at $h_{tc}=2.189$. However, we have found one stable branch of coexisting equilibrium when $h$ lies in $[0, h_{sn2}]$ with $h_{sn2}=0.236$. Now, we have found that the system exhibits a saddle-node bifurcation at $h=h_{sn1}=0.191$, from which two branches of interior points evolve. Among these two branches, one remains always stable and ultimately emerges with the predator-free state at $h=h_{tc}$ through a transcritical bifurcation. The other branch acts as a saddle point, which coincides with the first mentioned stable branch through another saddle-node bifurcation at $h=h_{sn2}$. So, a region comes into sight when $h$ lies in $[h_{sn1}, h_{sn2}]$ where we find a total of three branches of interior points, among which two are stable, and one is a saddle point. It indicates the occurrence of bistability in the system. The choice of initial population size plays a crucial role in this case. A trajectory tends to that equilibrium point from whose basin of attraction it is initiated. A small perturbation in the initial size can change the system dynamics and the trajectory can settle down to a different state from the state it was supposed to converge.  

\begin{figure}[htb!]
    \centering
     \centering
    \begin{subfigure}[t]{0.5\textwidth}
        \centering
        \includegraphics[width=\linewidth]{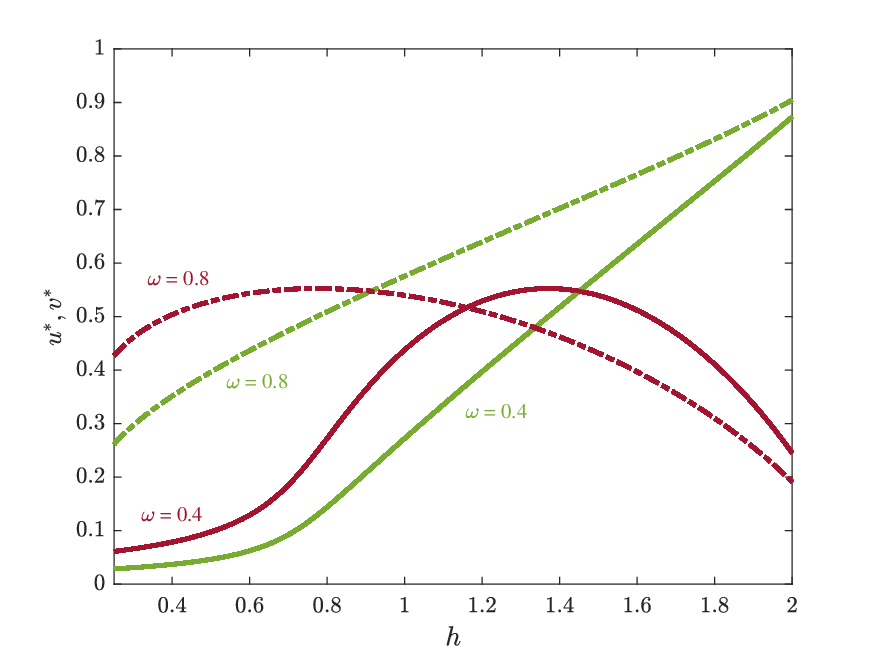}
        \caption{}\label{fig:4a}
    \end{subfigure}%
    ~ 
    \begin{subfigure}[t]{0.5\textwidth}
        \centering
        \includegraphics[width=\linewidth]{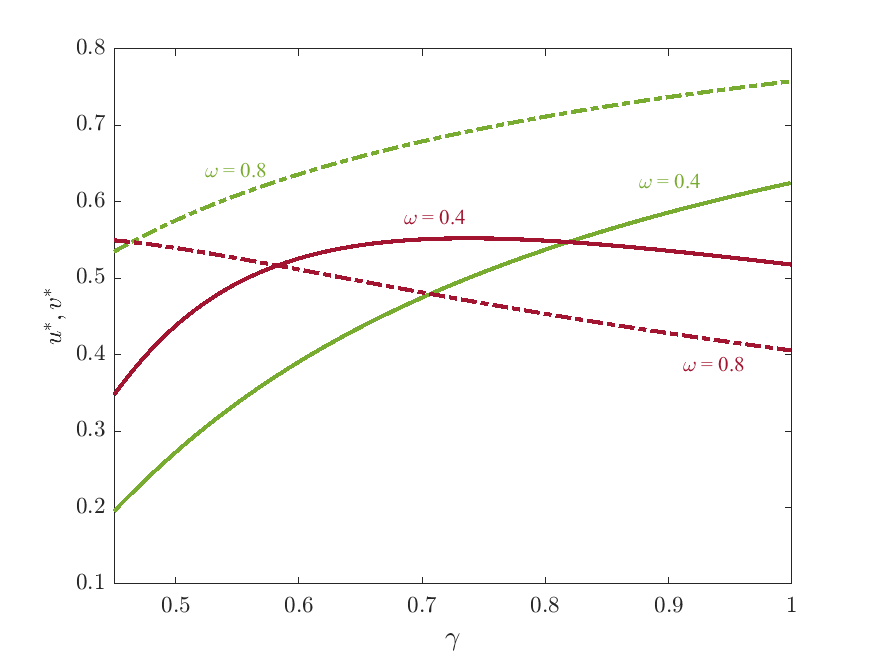}
        \caption{}\label{fig:4b}
    \end{subfigure}
    \caption{Change of coexisting state $(u^{*}, v^{*})$ with the change of (a) $h$ (the time required for the predator to handle a prey) and (b) $\gamma$ (predator interference) for different defence levels. The prey and predator biomass are represented by ({\color{GRN}\solidrule}) and ({\color{MRN}\solidrule}) colored graphs, respectively. Parameter values are given in Table \ref{Table:1}.} \label{fig:4}
\end{figure}

Figure \ref{fig:4a} depicts how the inducible defence affects the species' density if the prey handling time $(h)$ starts to increase. In addition, the steepness is also higher for $\omega=0.8$ than $\omega=0.4$. On the other hand, the predator biomass shows interesting dynamics. For a lower inducible defence level, the predator density rises more rapidly and reaches a peak before diminishing at a steep rate. It indicates if the prey has a lower defence level, the predator density increases intensely with increasing prey handling rate but it is more susceptible to a drip for a substantial rate. Here, the predator density increases initially with $h$ in the presence of strong inducible defence, but after a further increment, it starts to decline for a higher handling rate. It indicates if the prey has a stronger defence level, the predator's biomass wanes markedly for a higher prey handling rate. As we have considered the consumption rate with predator interference, hence, we have portrayed how it affects the species count in the presence of inducible defence. Figure \ref{fig:4b} shows that the prey biomass increases for a higher value of $\gamma$, but the increment rate is less steep for higher defence level $(\omega=0.8)$. This result shows that the predator count decreases monotonically with the increase of predator interference if the prey holds a stronger defence level. 


\subsection{Spatio-temporal dynamics}

This subsection illustrates the impact of diffusion in predator-prey interaction, and we have considered a bounded square domain as $\Omega=[0, L]\times[0, L]$ with $L=100$. For numerical simulations, the forward Euler method has been used for time integration with time increment $dt=0.005$, and the finite difference scheme has been used for the spatial derivatives with grid-spacing $dx=dy=0.5$. However, the qualitative results remain the same for small grid-spacing values. In addition, heterogeneous perturbations around the coexisting homogeneous steady-state are considered for the initial conditions, which is given by $u(x,y,0)=u^{*}+\epsilon\eta(x,y)$ and $v(x,y,0)=v^{*}+\epsilon\psi(x,y)$ with $\epsilon=10^{-3}$, where $\eta$ and $\psi$ are Gaussian white noise in space.


\subsection*{Effect of inducible defence in pattern formation in the absence of taxis}
As we have observed in the temporal system, if the defence level is chosen as the controlling parameter, then the system is unstable when $\omega<\omega_{H}$ and becomes stable while exceeding the threshold value. 
\begin{figure}[htb!]
    \centering
    \begin{subfigure}[t]{0.3\textwidth}
        \centering
        \includegraphics[width=\linewidth]{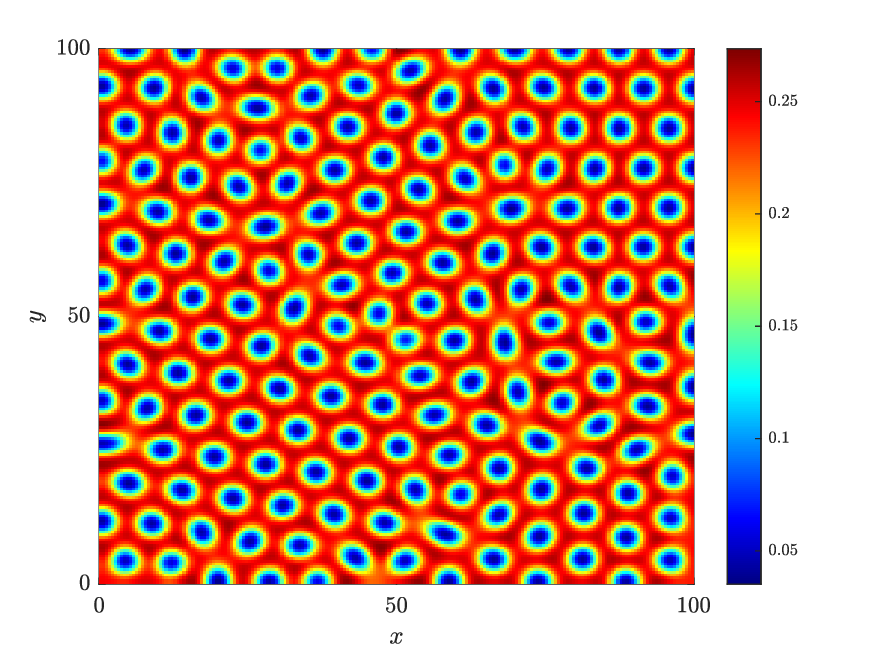}
        \caption{$(\omega, d_{1})=(0.27, 0.1)$}\label{fig:6a}
    \end{subfigure}
    ~ 
    \begin{subfigure}[t]{0.3\textwidth}
        \centering
        \includegraphics[width=\linewidth]{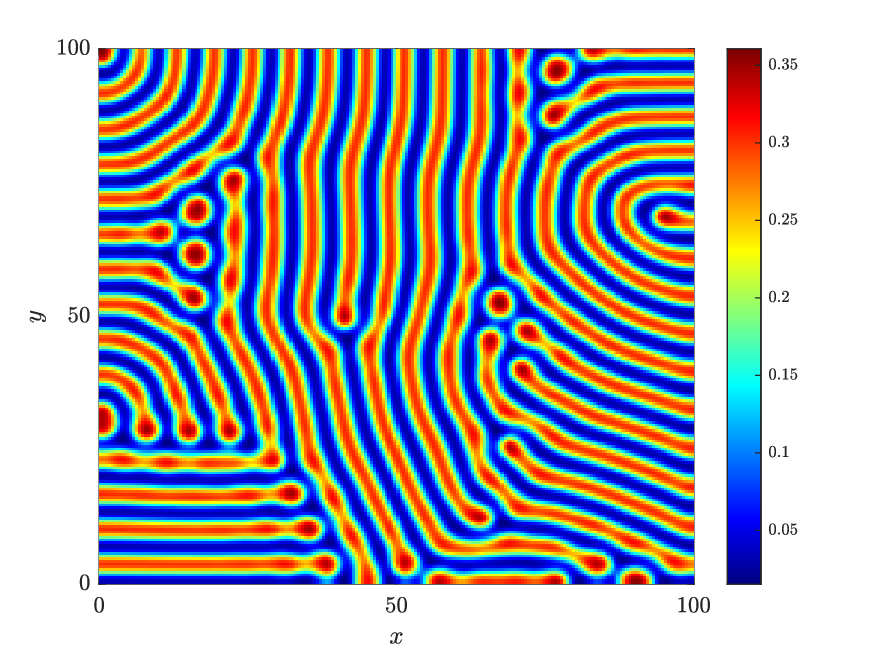}
        \caption{$(\omega, d_{1})=(0.15, 0.1)$}\label{fig:6b}
    \end{subfigure}
    ~ 
    \begin{subfigure}[t]{0.3\textwidth}
        \centering
        \includegraphics[width=\linewidth]{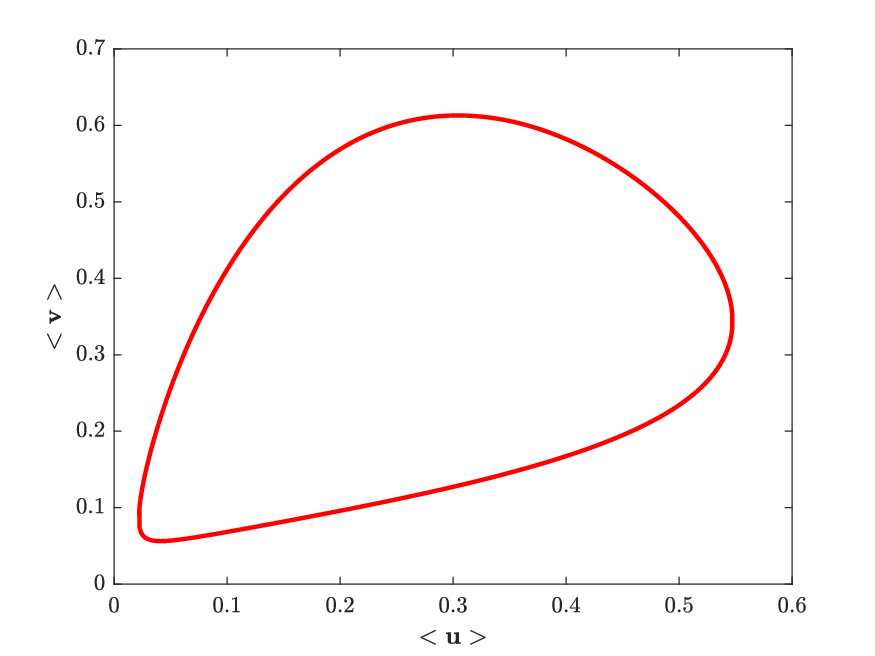}
        \caption{$(\omega, d_{1})=(0.15, 0.5)$}\label{fig:6c}
    \end{subfigure}
    \caption{Stationary patterns for the prey species $(u)$ when $(\omega, d_{1})$ is chosen from (a) Turing domain and (b) Turing-Hopf domain. (c) Spatial average of $u$ and $v$ for $t\in [25000, 50000]$ when $(\omega, d_{1})$ is chosen from Hopf domain.} \label{fig:6}
\end{figure}
The dynamical behaviours of the proposed spatio-temporal model are portrayed in Fig. \ref{fig:6} when the species disperse in a two-dimensional domain, and $(\omega, d_{1})$ are chosen from different domains. For $\omega=0.27$, the feasible interior equilibrium $E^{*}=(0.187, 0.336)$ is stable and the Turing bifurcation threshold $(d_{1c})$ is $0.121$. Stationary Turing patterns are observed in this case when $d_{1}<d_{1c}$ [see Fig. \ref{fig:6a}]. Moreover, at $\omega=0.15$ (i.e., $\omega<\omega_{H}$), the interior point $E^{*}$ becomes unstable, and $d_{1c}$ takes the value 0.231. When $\omega$ is chosen from the Hopf unstable domain, the system exhibits non-homogeneous stationary patterns or oscillatory solutions depending on the position of $d_{1}$ relative to the Turing curve. Specifically, non-homogeneous stationary patterns emerge when $d_{1}$ lies below the Turing curve, while oscillatory solutions appear when $d_{1}$ lies above it [see Figs. \ref{fig:6b} and \ref{fig:6c}]. 

\subsection*{Implementation of attraction-repulsion taxis in the spatio-temporal model}
Along with random diffusion of species, we have also considered the situation when the prey avoids the higher predator zone (chemorepulsion or specifically predator-taxis), and the predator moves towards the higher prey zone (chemoattraction or specifically prey-taxis). 
\begin{figure}[hbt!]
    \centering
    \begin{subfigure}[t]{0.32\textwidth}
        \centering
        \includegraphics[width=\linewidth]{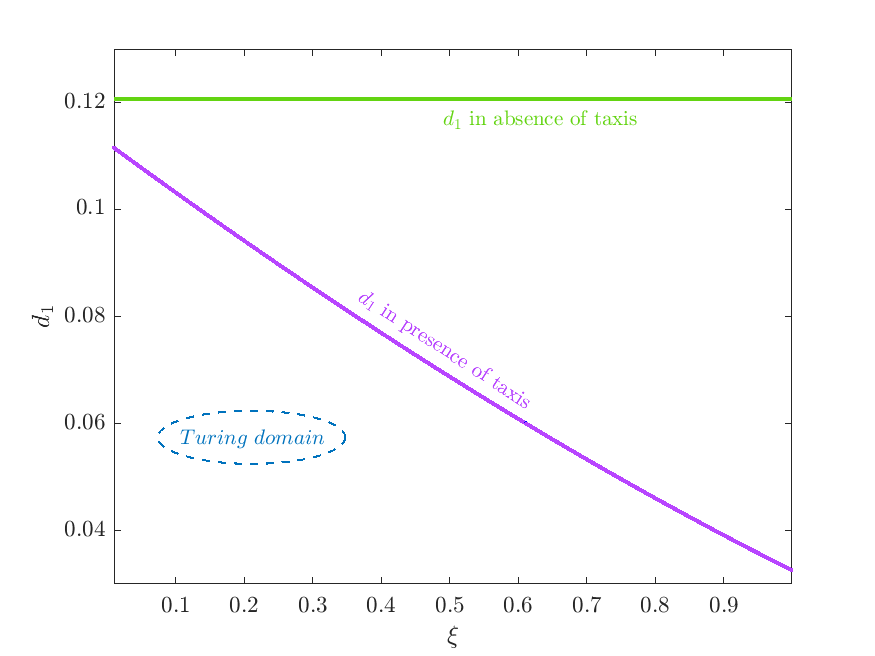}
        \caption{}\label{fig:7a}
    \end{subfigure}
    ~
    \begin{subfigure}[t]{0.32\textwidth}
        \centering
        \includegraphics[width=\linewidth]{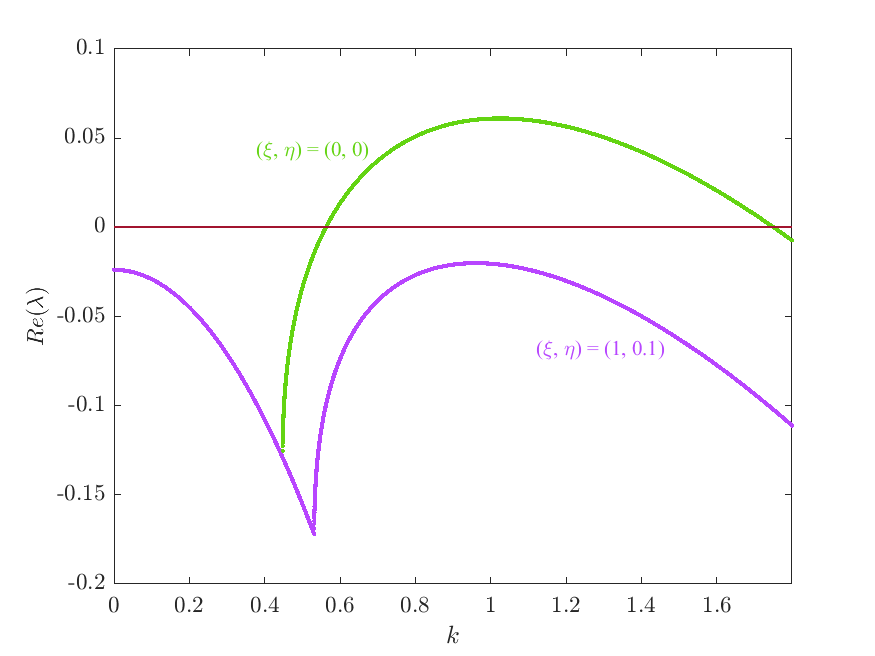}
        \caption{}\label{fig:7b}
    \end{subfigure}%
    ~
    \begin{subfigure}[t]{0.32\textwidth}
        \centering
        \includegraphics[width=\linewidth]{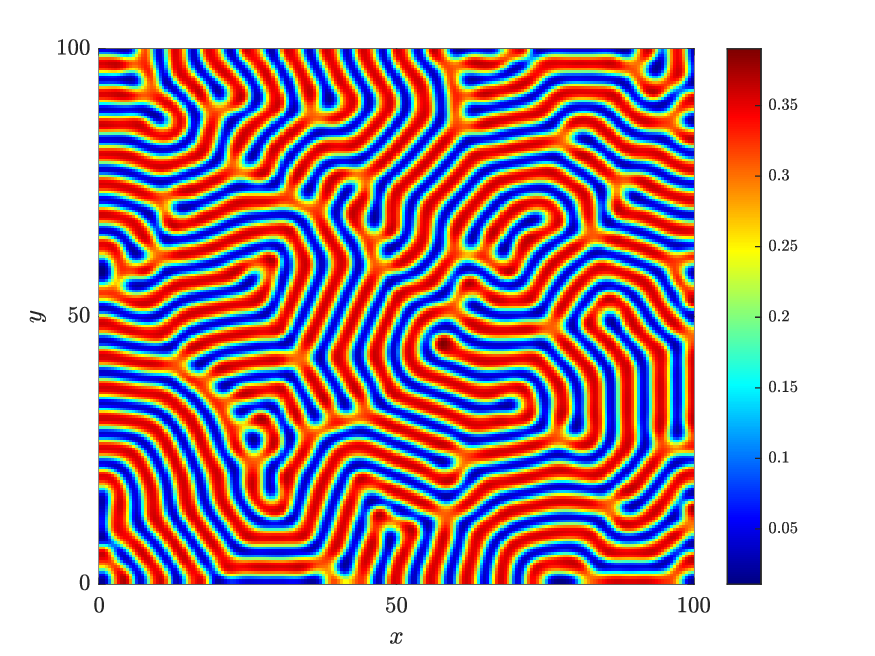}
        \caption{}\label{fig:7c}
    \end{subfigure}
\caption{(a) Critical diffusion coefficient $(d_{1c})$ in absence and in presence of $\xi$. (b) Plots of $\max(\mbox{Re}(\lambda))$ for $k$ in the presence as well as the absence of taxis for $d_{1}=0.05$. (c) Solutions for the prey species $(u)$ of the model (\ref{eq:diff2}). Parameter values are given in Table \ref{Table:1} and $(\omega, \xi, \eta)=(0.27, 1, 0.1)$.} \label{fig:7}
\end{figure}
In Fig. \ref{fig:7a}, it is depicted that the inclusion of taxis shrinks the Turing domain, lowering the chances of pattern formation of both species. For $\omega=0.27$, the Turing threshold $(d_{1c})$ is $0.121$, but when taxis are introduced $[(\xi, \eta)=(1,0.1)]$, the threshold becomes $0.033$. The Turing thresholds differ at $\xi=0$ because the system accounts for the directed movement of predators toward regions of higher prey density $(\eta=0.1)$. However, when $\eta=0$, the threshold values coincide at $\xi=0$. In this scenario, the prey diffusion threshold remains higher with increasing $\xi$ compared to the case where directed predator movement is present. Figures \ref{fig:7b} and \ref{fig:7c} demonstrate that for $d_{1}=0.05$, the spatio-temporal model produces a stationary (labyrinthine) pattern in the absence of taxis, but not in the presence of taxis.

\subsection*{ Impact of prey defence $(\omega)$ in the stabilization effect of taxis}
\begin{figure}[htb!]
    \centering
    \begin{subfigure}[t]{0.5\textwidth}
        \centering
        \includegraphics[width=\linewidth]{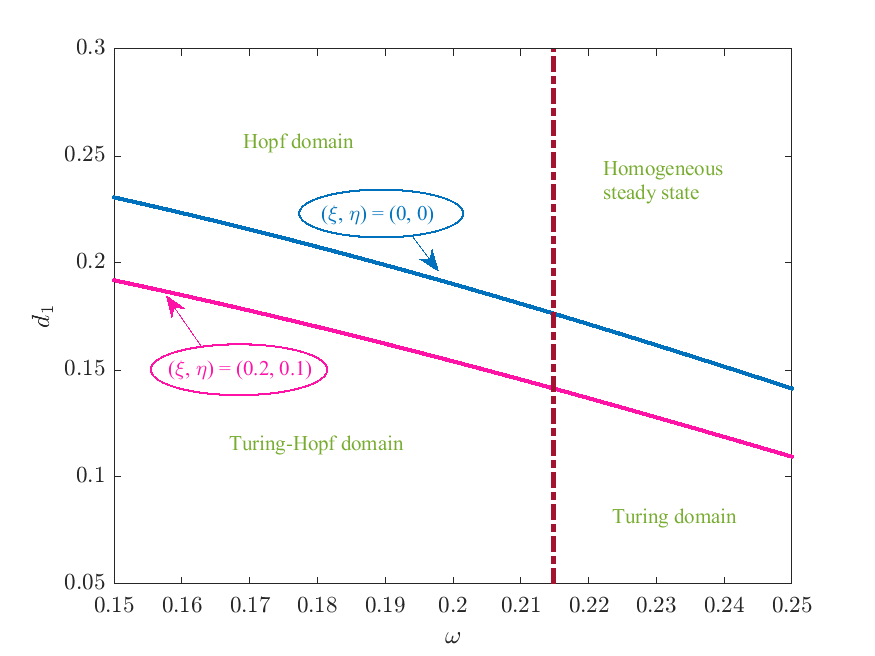}
        \caption{}\label{fig:8a}
    \end{subfigure}%
    ~
    \begin{subfigure}[t]{0.5\textwidth}
        \centering
        \includegraphics[width=\linewidth]{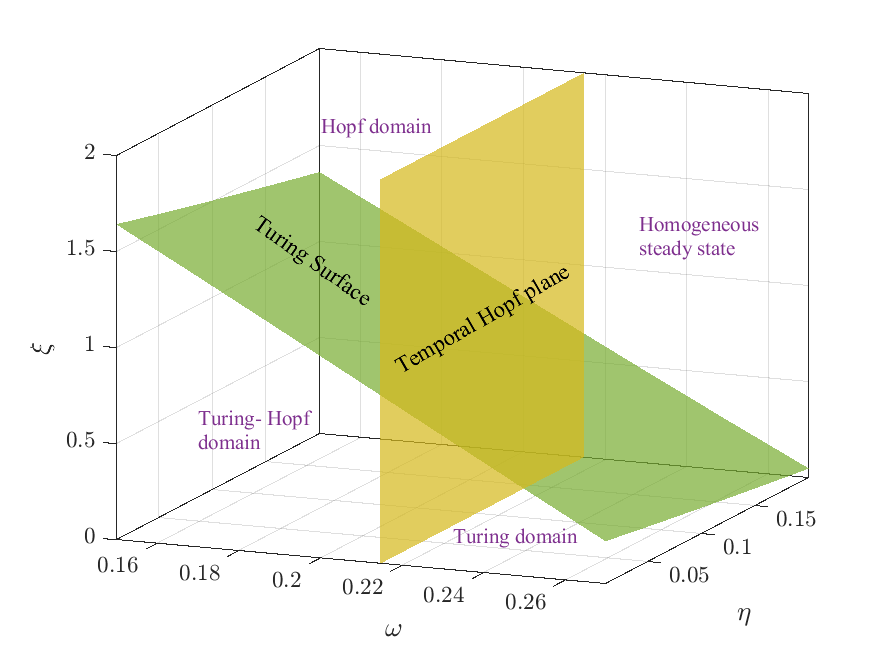}
        \caption{}\label{fig:8b}
    \end{subfigure}
\\
    \begin{subfigure}[t]{0.3\textwidth}
        \centering
        \includegraphics[width=\linewidth]{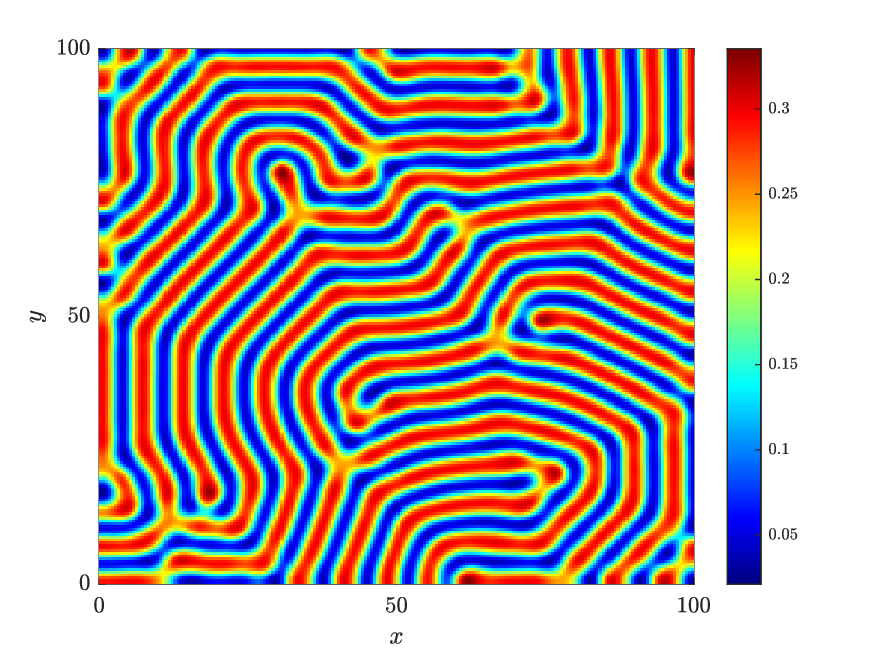}
        \caption{}\label{fig:8c}
    \end{subfigure}
    ~ 
    \begin{subfigure}[t]{0.3\textwidth}
        \centering
        \includegraphics[width=\linewidth]{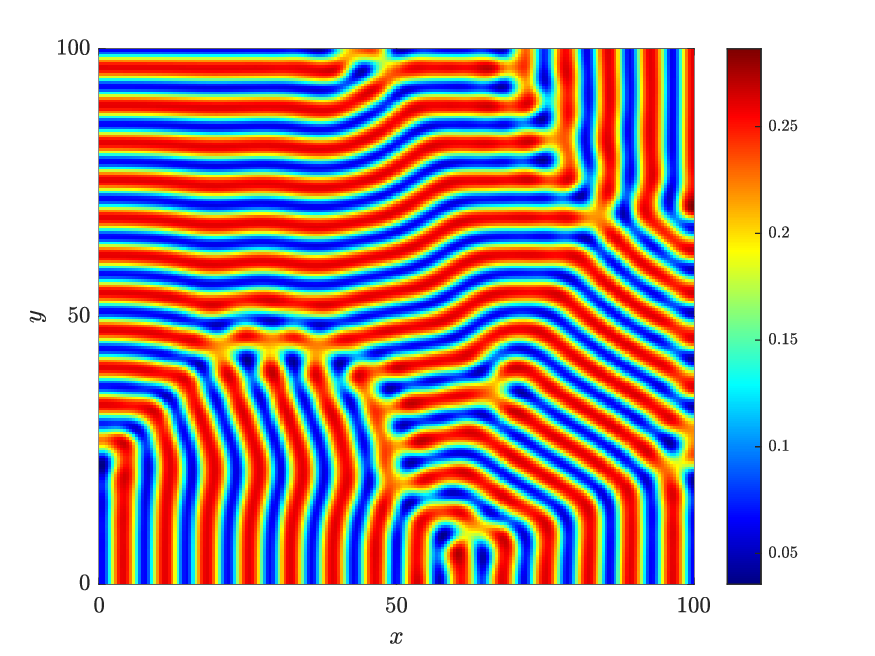}
        \caption{}\label{fig:8d}
    \end{subfigure}
    ~ 
    \begin{subfigure}[t]{0.3\textwidth}
        \centering
        \includegraphics[width=\linewidth]{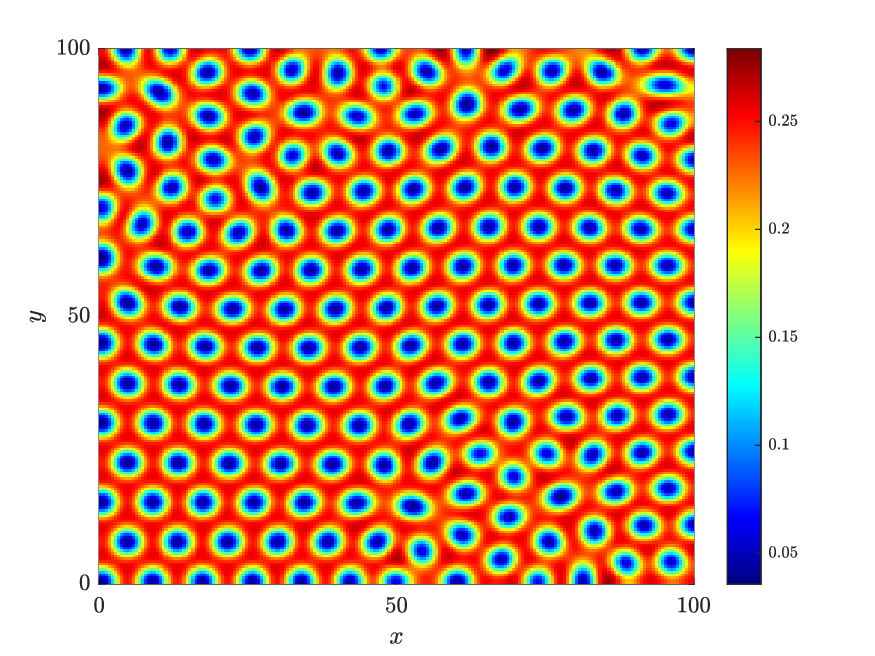}
        \caption{}\label{fig:8e}
    \end{subfigure}
    \caption{(a) Temporal-Hopf ({\color{darkmar}\protect\dashdotrule}) and Turing bifurcation curves in the $\omega$-$d_{1}$ plane in the presence of taxis ({\color{mag}\solidrule}) as well as in the absence of taxis ({\color{fadblu}\solidrule}). (b) Occurrence of Turning surface and Temporal-Hopf plane in $\omega$-$\eta$-$\xi$ domain. (c)-(e) Occurrence of non-homogeneous stationary pattern in the presence and absence of taxis for $\omega=0.22$, but for $\omega=0.27$, patterns occur in the absence of taxis only. 
    Parameter values are mentioned in Table \ref{Table:1} and $(d_{1}, \xi, \eta)=(0.1, 0.2, 0.1)$.} \label{fig:8}
\end{figure}

We have discussed how the inducible defence impacts the spatio-temporal dynamics when taxis are incorporated into the system. Figure \ref{fig:8a} shows that increasing defence reduces species colonization scope by shrinking the Turing domain. It indicates that an elevated defence level in prey species decreases the prey diffusion threshold $d_{1}$ for the emergence of Turing patterns, thereby constricting the parametric region conducive to patch formation. So, a higher level of defence in prey contracts the chances of patch formation. 
Inducible defences often make prey harder to catch or less appealing to predators. A reduced Turing domain implies that these defences confine the places where predators may readily capture prey, potentially enhancing prey survival rates. Predator-prey interactions may become more localized when the Turing domain is shrunk. Turing patterns help to maintain ecosystem diversity and stability by forming niches and sustaining a wide range of species. A smaller area might diminish spatial complexity, perhaps resulting in a more homogeneous environment with distinct consequences for biodiversity. Moreover, the Turing surface is plotted in the presence of taxis, where the stabilizing effect of taxis is shown for increasing inducible defence [see Fig. \ref{fig:8b}]. For $d_{1}=0.1$, if we choose $\omega=0.22(>\omega_{H})$, we get a stationary Turing pattern in the system, in the presence as well as in the absence of prey- and predator-taxis [see Figs. \ref{fig:8c}, \ref{fig:8d}]. Then again, for $\omega=0.27$, stationary patterns are observed only in the absence of taxis [see Fig. \ref{fig:8e}]. This figure illustrates that when both prey- and predator-taxis are implemented into a model where prey species show inducible defence against their predators, the scope of stationary patch formation diminishes significantly. As the prey moves away from the predators by adopting a defence strategy and the predators move towards prey, the spatially stable patterns are hard to sustain. From a biological point of view, the amalgamation of adaptive movement and increased defences obstruct the formation of patches in the ecosystem. Instead of steady, predictable patches, the prey and predator populations may exhibit a homogeneous spread in this case.

\subsection*{Particular case: $\eta=0$ but $\xi\neq 0$}
Let us consider a particular scenario where the prey population not only moves from higher to lower concentration (diffusion) but avoids the places with higher predator density (predator-taxis). But, the predator's searching strategy is merely restricted to random search $(\eta=0)$. 
\begin{figure}[htb!]
    \centering
    \begin{subfigure}[t]{0.3\textwidth}
        \centering
        \includegraphics[width=\linewidth]{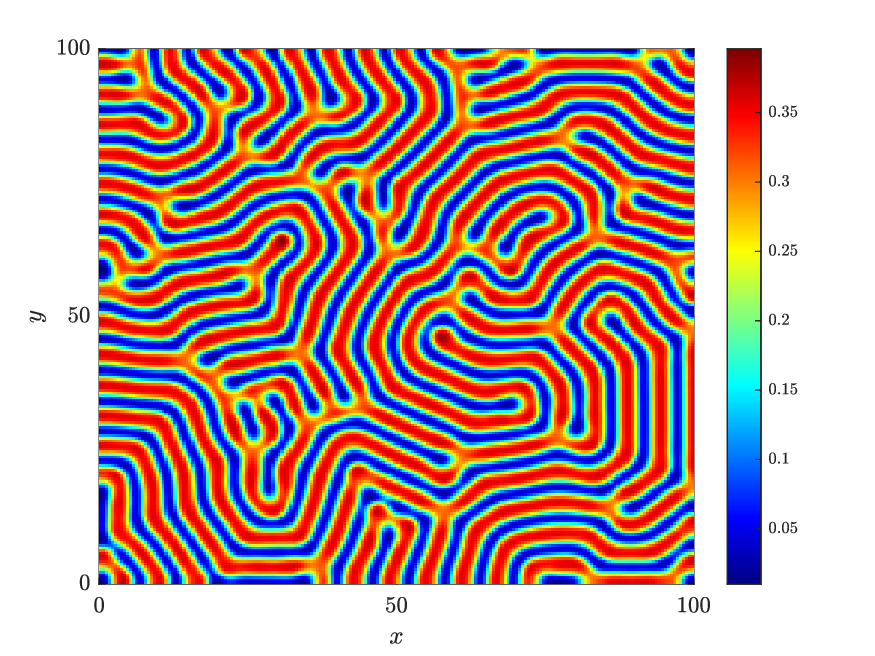}
        \caption{$(\omega,\xi)=(0.27, 0.01\xi_{c})$}\label{fig:9a}
    \end{subfigure}
    ~ 
    \begin{subfigure}[t]{0.3\textwidth}
        \centering
        \includegraphics[width=\linewidth]{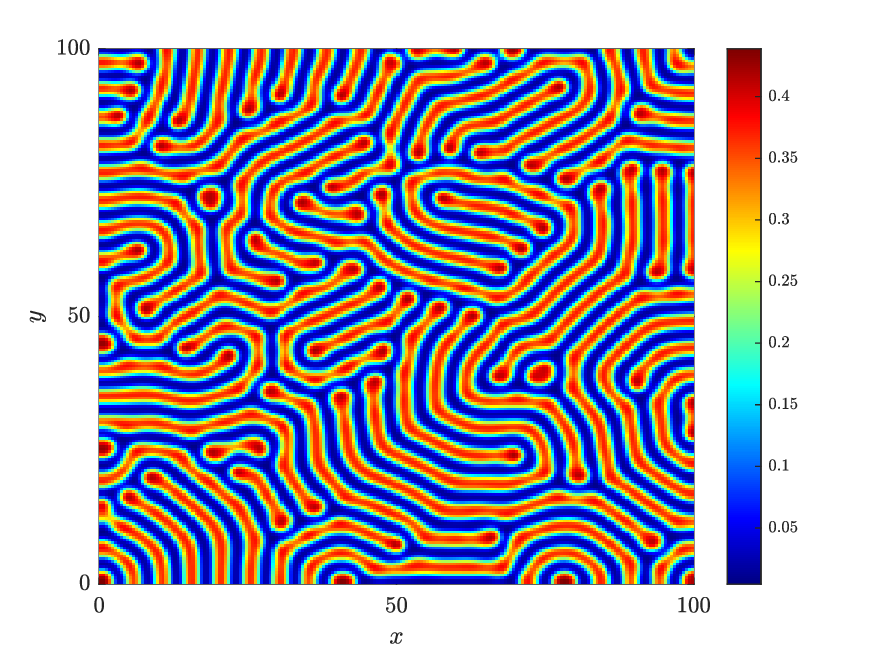}
        \caption{$(\omega,\xi)=(0.15, 0.01\xi_{c})$}\label{fig:9b}
    \end{subfigure}
    ~ 
    \begin{subfigure}[t]{0.3\textwidth}
        \centering
        \includegraphics[width=\linewidth]{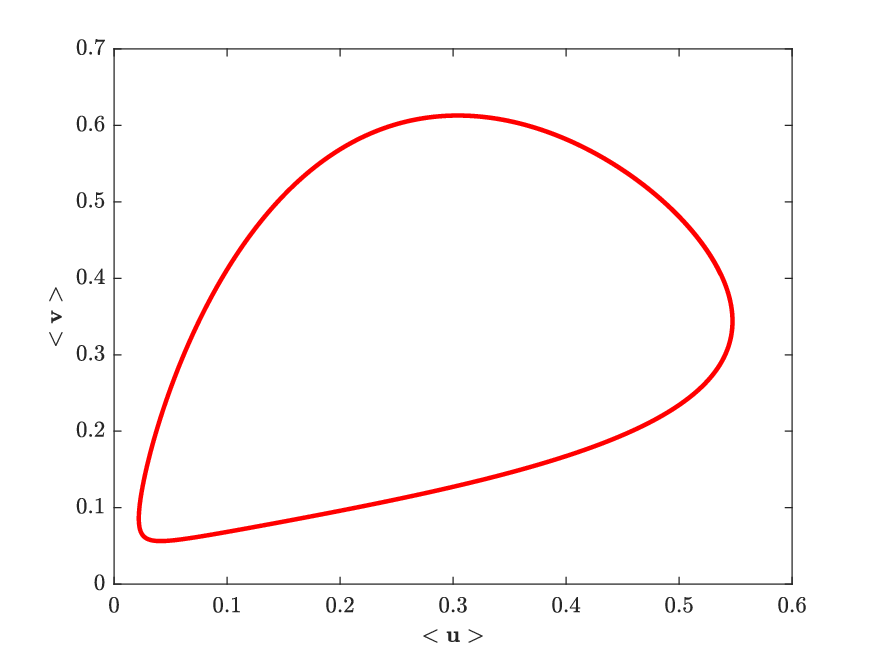}
        \caption{$(\omega,\xi)=(0.15, 3\xi_{c})$}\label{fig:9c}
    \end{subfigure}
    \caption{(a) -- (b) Solutions for the prey species $(u)$ when $(\omega,\xi)$ is chosen from different domain and $(\eta, d_{1})=(0, 0.05)$. (c) Spatial average of $u$ and $v$ in Hopf domain for $t\in [40000, 50000]$. Parameter values are given in Table \ref{Table:1}.} \label{fig:9}
\end{figure}
Choosing $d_{1}$ as $0.05$, the critical sensitivity coefficient for predator-taxis $(\xi_c)$ is found to be $0.251$ for $\omega=0.15$ and $0.089$ for $\omega=0.27$. In Fig. \ref{fig:9}, the dynamic behaviour is shown when $(\omega, \xi)$ is chosen from Turing [Fig. \ref{fig:9a}], Turing-Hopf [Fig. \ref{fig:9b}], and Hopf domains [Fig. \ref{fig:9c}]. 

\subsection*{ Impact of predator interference $(\gamma)$ in the stabilization effect of taxis}

Not only the inducible defence, but one of the focal points is to explore the role of predator interference in pattern formation in the presence and absence of taxis. Not only the inducible defence, but one of the focal points is to explore the role of predator interference in pattern formation in the presence and absence of taxis.
\begin{figure}[htb!]
    \centering
    \begin{subfigure}[t]{0.245\textwidth}
        \centering
        \includegraphics[width=\linewidth]{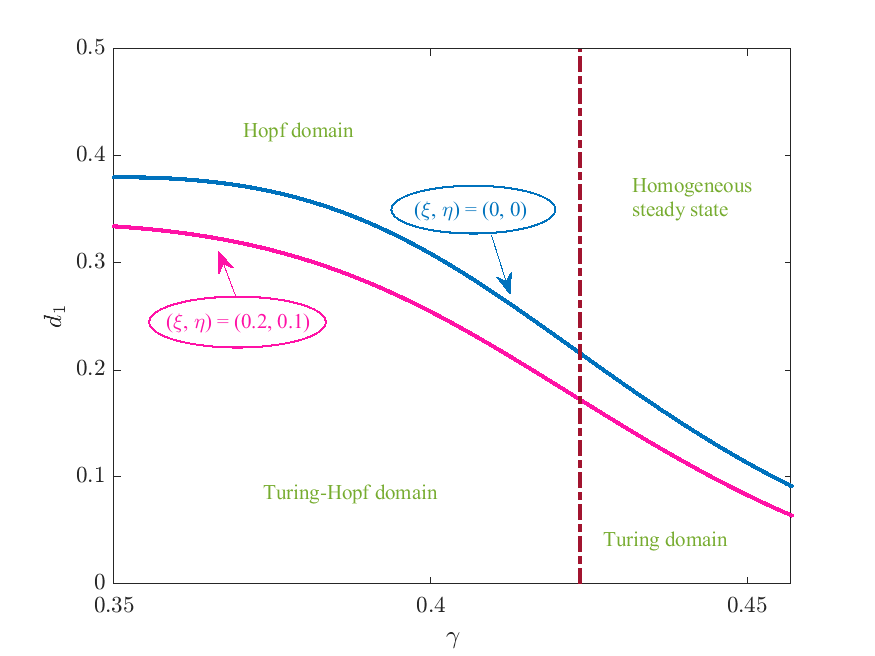}
        \caption{}\label{fig:10a}
    \end{subfigure}%
    \begin{subfigure}[t]{0.245\textwidth}
        \centering
        \includegraphics[width=\linewidth]{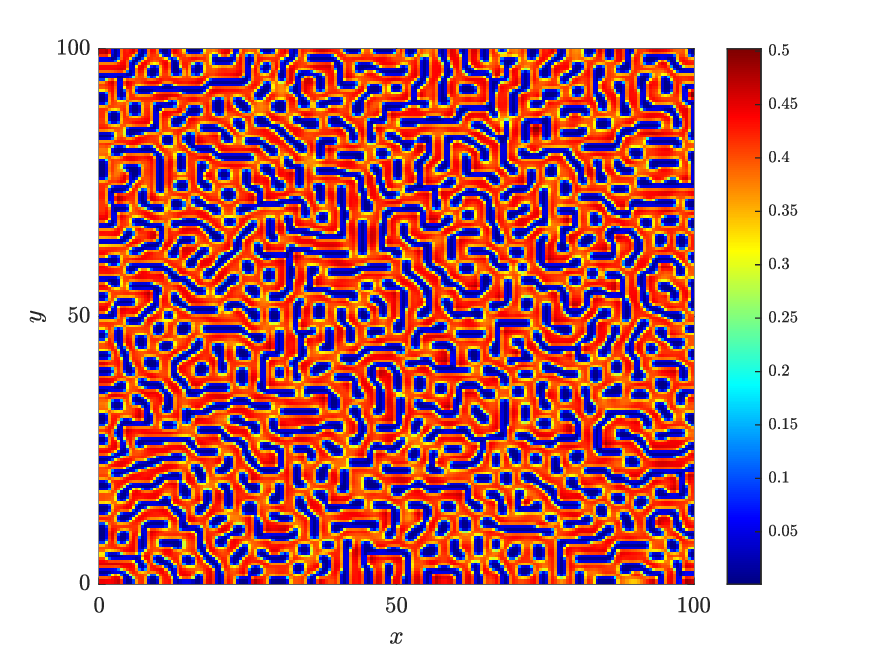}
        \caption{}\label{fig:10b}
    \end{subfigure}
    \begin{subfigure}[t]{0.245\textwidth}
        \centering
        \includegraphics[width=\linewidth]{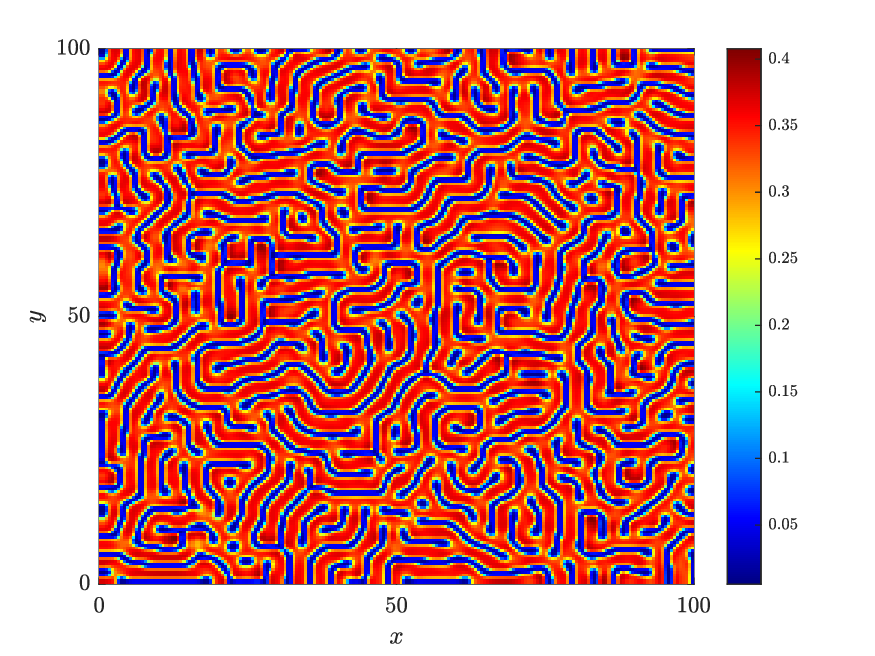}
        \caption{}\label{fig:10c}
    \end{subfigure}
    \begin{subfigure}[t]{0.245\textwidth}
        \centering
        \includegraphics[width=\linewidth]{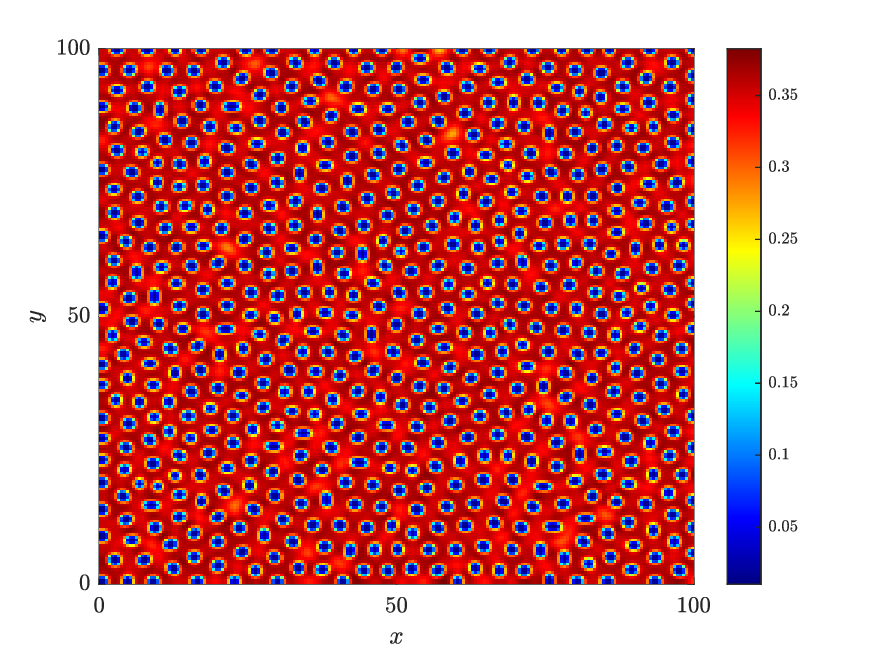}
        \caption{}\label{fig:10d}
   \end{subfigure}
   \caption{(a) Temporal-Hopf ({\color{darkmar}\protect\dashdotrule}) and Turing bifurcation curves in the $\gamma$-$d_{1}$ plane in the presence of taxis ({\color{mag}\solidrule}) as well as in the absence of taxis ({\color{fadblu}\solidrule}). (b)-(d) Occurrence of non-homogeneous stationary pattern in the presence and absence of taxis for $\gamma=0.45$, but for $\gamma=0.50$, patterns occur in the absence of taxis only. 
   Parameter values are mentioned in Table \ref{Table:1} and $(\omega, d_{1}, \xi, \eta)=(0.4, 0.01, 0.2, 0.1)$.} \label{fig:10}
\end{figure}
Figure \ref{fig:10a} depicts how the increase in interference rate lowers the prey diffusion coefficient $(d_{1})$ contracting the Turing domain. It signifies that higher predator interference can dampen predator efficacy, leading to stable or patterned distribution instead of chaotic fluctuations. The Turing threshold $(d_{1c})$ exhibits a notable reduction when $(\xi, \eta)$ are assigned non-negative values, such as $(\xi, \eta)=(0.2, 0.1)$. This behaviour highlights a natural inclination of the species to settle into a homogeneous state. From an ecological perspective, in higher predator interference, the predator finds it difficult to hunt effectively, resulting in a more homogeneous distribution of both species. In particular, the directed movement of a species towards (or opposite of) another species, implemented in the system, cannot create considerable variation in population density, inducing homogeneous species distribution. Figures \ref{fig:10b} and \ref{fig:10c} show that the spatio-temporal system produces a non-homogeneous stationary pattern when $\gamma=0.45$ (Hopf stable domain) and $d_{1}=0.01$ in the presence as well as in the absence of taxis, respectively. But, when $\gamma$ is chosen as $0.50$, the Turing patterns are observed only when no directed movement of species is induced [see Fig. \ref{fig:10d}]. 

\subsection*{Role of reverse taxis}
Now, repulsive predator-taxis and attractive prey-taxis are phenomena in which prey species avoid higher predator zones, and predators prefer to move toward higher prey zones, respectively. Consider a situation with attractive predator-taxis and repulsive prey-taxis $(\xi<0, \ \eta<0)$. Attractive predator-taxis implies the tendency of prey moving toward the predator species. This may occur when the prey population has absolute numerical dominance while the predator is small. In addition, the prey moves toward the predators for collective defence against the predator attack or to rescue their companions under attack by predators. When this defence becomes strong enough, the predators start to avoid the high prey density areas, resulting in the occurrence of repulsive prey-taxis $(\eta<0)$. 
\begin{figure}[htb!]
    \centering
    \begin{subfigure}[t]{0.3\textwidth}
        \centering
        \includegraphics[width=\linewidth]{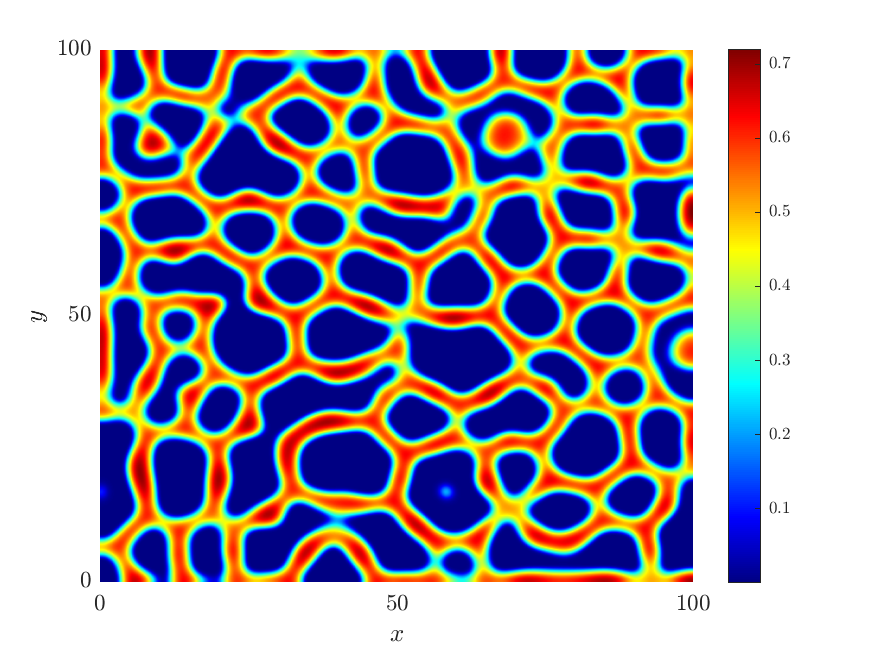}
        \caption{}\label{fig:11a}
    \end{subfigure}
    ~ 
    \begin{subfigure}[t]{0.3\textwidth}
        \centering
        \includegraphics[width=\linewidth]{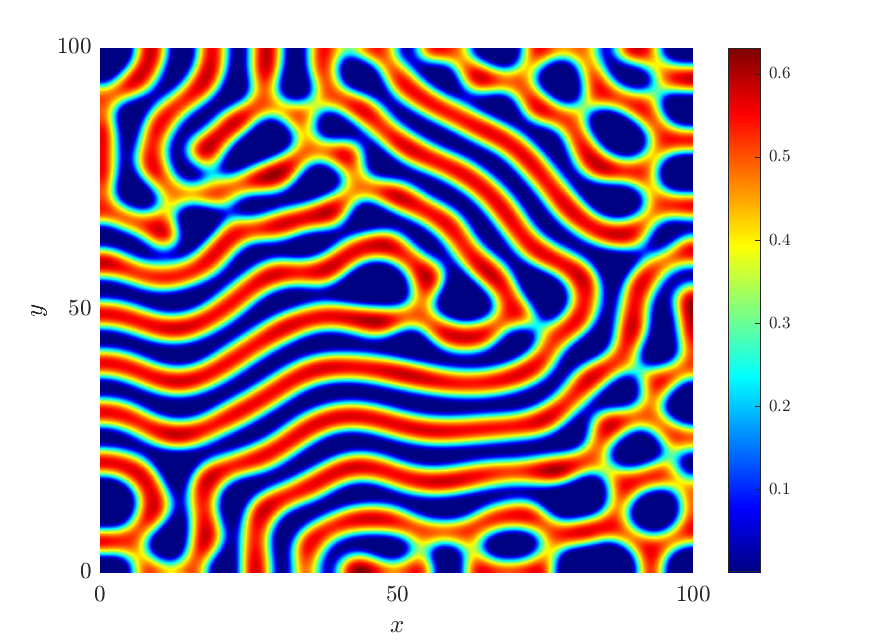}
        \caption{}\label{fig:11b}
    \end{subfigure}
    ~ 
    \begin{subfigure}[t]{0.3\textwidth}
        \centering
        \includegraphics[width=\linewidth]{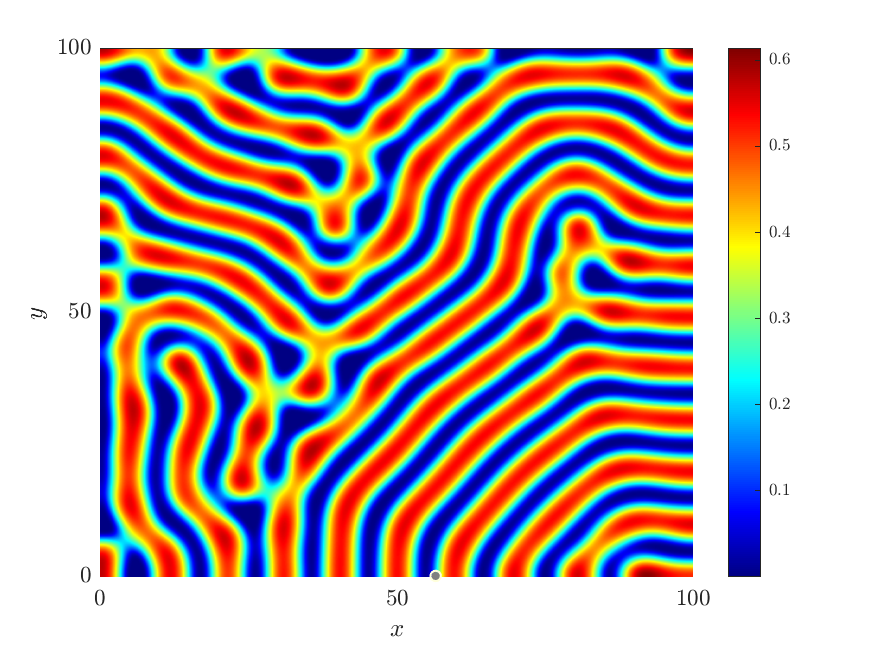}
        \caption{}\label{fig:11c}
    \end{subfigure}
    \caption{Solutions for the prey species $(u)$ of the model (\ref{eq:diff3}) when reverse taxis are applied by choosing $\xi=-2$ along with (a) $\eta= -1$, (b) $\eta=-2$ and (c) $\eta=-3$. Parameter values are given in Table \ref{Table:1} and $\omega=0.27$.} \label{fig:11}
\end{figure}
In Fig. \ref{fig:11}, the dynamic scenario is depicted for this situation when $(\omega, d_{1})=(0.27, 0.05)$. We have chosen $dx=dy=0.2$ and $dt=0.0005$ for this figure. The scenario reveals the occurrence of non-homogeneous stationary patterns when $(\xi, \eta)=(-2, -1)$ [Fig. \ref{fig:11a}], $(\xi, \eta)=(-2, -2)$ [Fig. \ref{fig:11b}] and $(\xi, \eta)=(-2, -3)$ [Fig. \ref{fig:11b}]. Not only that, the figures show that the system produces labyrinthine patterns when $\xi>\eta$, but for $\xi<\eta$, it does not. 

Overall, the numerical simulations yield numerous critical insights into the proposed predator-prey system with inducible prey defence. We have demonstrated that predator numbers continue to climb while defence is minimal, but when defence rises, the prey population grows significantly, resulting in a fall in predator numbers. The temporal study reveals that inducible defence has a stabilizing influence on population counts. Furthermore, in the spatio-temporal model, defence lowers the size of the Turing domain, preventing spatial pattern formation. However, when a nonlocal term is included, the Turing domain gets bigger, allowing for complex spatial patterns. Depending on the value of the predator's diffusion coefficient in the presence of the nonlocal term, different outcomes, such as stationary, non-stationary, and oscillatory patterns, can emerge in the model system.

\section{Conclusions} \label{sec:7}

Inducible defences, a form of phenotypic plasticity, have the ability to significantly influence direct interactions within ecological communities, generating trait-mediated indirect effects \cite{wootton2002indirect, werner2003review}. These defences arise when prey exhibit adaptive behavioural, morphological, or physiological traits in response to their predators, effectually minimizing direct encounters with predators. However, such defences often come with associated costs- either through a reduction in prey growth rates (metabolic costs) or by impairing prey-resource interactions (feeding costs). Therefore, by changing the dynamics of interactions, inducible defences can have cascade trait-mediated impacts on prey, predators, and the prey's resources \cite{wootton1993indirect}. An evolutionary ecological theory posits that inducible defences are vouched for over constitutive ones when these defensive traits impose exceptional costs on prey \cite{harvell1990ecology, harvell1999inducible}. There are various predator-prey systems have been explored to elucidate the implications of inducible defences at both populations as well as community levels \cite{abrams1996invulnerable, ramos2000relating, ramos2002population, ramos2003population, kopp2006dynamic}. However, the focus of our study is to look over the impact of inducible defences on the dynamic behaviour of predator-prey interactions, particularly in the context of predator interference and the coupled effects of repulsive and attractive taxis. \par

In this work, we have developed a predator-prey interaction that includes psychological stress in the prey species induced by the prey's inducible defence strategy. This defensive mechanism in prey is included in this model that allows prey to respond to predators by developing adaptive features (higher speed or camouflage), which lowers the pressure of predation. The main intention here is to elucidate the importance of this factor in the dynamic behaviour of the model, as we have assumed that the growth rate of prey is significantly affected due to the strategy. Moreover, we have considered the Beddington-DeAngelis functional response while formulating the model. Because both the prey and predator densities influence the predator-prey interaction with this response, a more realistic saturation effect is possible when prey availability is higher. The incorporation of the defence has significantly increased the prey count in the system. Not only that, this defence is proven to be a stabilizing component for the system as increasing the level of defence helps to coexist the species in the environment as a stable steady-state. Moreover, it is shown that in the presence of a stronger defence level, the predator count still declines monotonically, even for lower predator interference. The handling rate of predator $(h)$ also plays a significant role in the system when inducible defence is there [see Fig. \ref{fig:4}]. \par

Instead of analyzing the temporal model only, we intend to focus on the involvement of species diffusion in the model. The species are assumed to move in a two-dimensional bounded domain. The analysis reveals that the diffusion coefficient for the prey species starts to decrease with increasing inducible defence level $(\omega)$ [see Fig. \ref{fig:8a}]. It indicates that the prey, because of adopting inducible defence, will ignore moving in the mentioned direction. In fact, the increase in $\omega$ shrinks the Turing domain, reducing the chances of non-homogeneous pattern formation. As the species are not always homogeneously distributed over a domain, this shrinkage may not be proven very favourable for persistence. The numerical simulation shows that the system exhibits cold spots when $\omega$ is chosen from the Hopf stable domain and $d_{1}<d_{1c}$ (Turing domain), but a labyrinthine pattern when chosen from the Hopf unstable domain (Turing-Hopf domain). \par

Furthermore, a spatial dynamic is added by directed diffusion for both species, named prey- and predator-taxis, in which prey tries to flee while predators aggressively approach. The repulsive predator-taxis $(\xi>0)$ and attractive prey-taxis $(\eta>0)$ have been chosen to perform the analysis here. It is observed through numerical simulation that the taxis effect tends to improve the spatial structure and inhibit runaway growth or decay of the species, and this can stabilize predator-prey dynamics by diminishing the Turing region. Additionally, it is also depicted that the prey diffusion coefficient $(d_{1})$ declines more in the presence of taxis when either of inducible defence level or predator interference rate increases in the system [see Figs. \ref{fig:8a}, \ref{fig:10a}]. This indicates that the inclusion of taxis can mitigate the destabilizing effects of diffusion, inducible defence, or even predator interference, resulting in complex spatial patterns that endure over time. As a special case, we have given a numerical illustration of the model with attractive predator-taxis $(\xi<0)$ and repulsive prey-taxis $(\eta<0)$ where non-homogeneous patterns are observed. \par

There are two ways in which inducible defences might help to stabilize the food chain. First, protective characteristics decrease the functional response, which hinders targeted predator-prey interaction. Additionally, the costs either weaken the prey-resource connection (feeding costs) or slow down the prey's rate of growth (metabolic costs). The weakening of trophic links is one of the stabilizing mechanisms in food webs, according to McCann et al. \cite{mccann1998weak}. On the other hand, predator abundance causes a decrease in contact intensity by creating a negative feedback loop \cite{kopp2006dynamic}. This self-regulating negative feedback process is generally acknowledged as an effective stabilizing component in population systems \cite{berryman2020principles, dambacher2007understanding}. As a result, interactions may be compromised by trait-mediated indirect effects in general and ID in particular, which would dampen oscillations and increase community stability. \par

Even with its complex dynamics, the proposed system can yet be enhanced in future work. The prey species' growth is influenced by their inducible defence is one of the prime considerations that is imposed by the system's design. However, in their natural environment, the predator species may employ other hunting methods as a counter-defence technique. To make the situation more realistic, it will be helpful to take into account the role that various functional responses in the growth of both the predator and the prey. Additionally, the carryover effect can be implemented in any predator-prey interaction in ecological systems where the past experiences and background of a species affect its present behaviour. The inducible defence of prey is one kind of phenotypic plasticity that modifies the direct interactions between various members of an ecological community and results in trait-mediated indirect effects. As so, it may affect several generations instead of simply one. It implies that the carryover effect may be incorporated by the prey species in the model. Furthermore, white Gaussian noise may be used to add ambient stochasticity into the system, providing even more realistic assumptions in the present framework.

\subsection{Acknowledgements}
The authors are grateful to the NSERC and the CRC Program for their support. RM is also acknowledging the support of the BERC 2022–2025 program and the Spanish Ministry of Science, Innovation and Universities through the Agencia Estatal de Investigacion (AEI) BCAM Severo Ochoa excellence accreditation SEV-2017–0718. This research was enabled in part by support provided by SHARCNET (\url{www.sharcnet.ca}) and Digital Research Alliance of Canada (\url{www.alliancecan.ca}).

\subsection{Data Availability Statement}
The data used to support the findings of the study are available within the article.

\subsection{Conflict of Interest}
This work does not have any conflict of interest.

\bibliographystyle{elsarticle-num} 
\bibliography{P_References}

\end{document}